\documentstyle[amssymb,epsfig]{mn2e}

\title[The Initial Mass Function of the Galactic Halo Globular Cluster System]
{New light on the Initial Mass Function of the Galactic Halo Globular Clusters}
\author[G.~Parmentier \& G.~Gilmore]{Genevi\`eve Parmentier 
\thanks{E-mail: gparm@ast.cam.ac.uk} \& Gerard Gilmore \\
Institute of Astronomy, University of Cambridge, 
Madingley Road, Cambridge CB3 0HA, United Kingdom}
\begin{document}

\date{Accepted .... Received ... ; in original form ...}

\pagerange{\pageref{firstpage}--\pageref{lastpage}} \pubyear{2004}

\maketitle

\label{firstpage}
 
\begin{abstract}
We present new constraints on the initial mass spectrum of the
Galactic Old Halo globular clusters.  This has remained
poorly-known so far, as both an initial power-law mass spectrum and an
initial lognormal mass function could have evolved into the presently
observed  
globular cluster mass distribution, making the initial contribution 
of now lost low-mass objects ill-determined.  Our approach consists 
in comparing the evolution with time of both the radial mass density 
profile and the number density profile of the globular cluster system.  
Using the analytical expression established by Vesperini \& Heggie 
for the temporal evolution of the mass of a globular cluster on a 
circular orbit in a stable Galactic potential, we evolve the mass and number
density profiles of many putative globular cluster systems, each
starting with different initial cluster mass spectrum and initial 
cluster space-density.  We then compare the modelled profiles with 
those of the Old Halo cluster system in order to investigate which
system(s) provide(s) the best consistency with the data.  
Specifically, we build on the following points: (1)~the presently
observed mass density profile and number density profile of the Old
Halo cluster system show the same shape, (2)~assuming that globular 
clusters were initially distributed the same way in mass at all 
galactocentric distances, the mass and number density profiles had 
also the same shape initially, (3)~according  to our simulations, 
the mass density profile remains well-preserved, irrespective of 
the initial cluster mass spectrum, while the temporal evolution of 
its number counterpart depends sensitively on initial conditions.  
We show that to obtain a mass density profile and a number density
profile which are identical in shape, both initially and after a
Hubble-time of evolution, the globular cluster system must have been
depleted in low-mass objects from the beginning.  We deduce
that the initial distribution in mass of the globular
clusters was either a lognormal mass function, similar to
that today, or a power-law mass spectrum with a slope $\simeq -2$ and
truncated at large mass, say, around $10^5$M$_{\odot}$.  In contrast,
a power-law mass spectrum with a similar slope but extending down to low
cluster mass (i.e., a few thousand solar masses) seems to be ruled out.  
\end{abstract}


\section{Introduction}

Globular clusters are the oldest bound stellar systems
in our Galaxy.  Their study provides therefore valuable information 
about early Galactic evolution.  In this respect, a major problem 
is that we do not know whether what we presently observe is still 
representative of the initial conditions and, thus, a fossil imprint
of the formation process, or whether the initial conditions have been 
wiped out by a 13\,Gyr long evolution within the tidal fields of
what is now the Milky Way.  
Modelling the dynamical evolution of the Galactic globular cluster 
system is thus of great interest as it helps us to go back
in time to the earliest stages of the cluster system and to disentangle
the formation and evolutionary fingerprints (see, e.g., 
Okazaki \& Tosa 1995, Baumgardt 1998, Vesperini 1998, Fall \& Zhang
2001).  In spite of numerous efforts however, the shape of the initial 
distribution in mass of the halo globular clusters has remained 
ill-determined so far. \\

In our Galaxy, the mass function of the halo clusters 
\footnote{In what follows, we adopt the nomenclature of 
McLaughlin \& Pudritz (1996).  We call luminosity/mass {\sl spectrum} 
the number of objects per {\sl linear} luminosity/mass interval, 
${\rm d}N/{\rm d}L$ or ${\rm d}N/{\rm d}m$,
while we refer to the luminosity/mass {\sl function} to describe
the number of objects per {\sl logarithmic} luminosity/mass interval, 
${\rm d}N/{\rm dlog}~L$ or ${\rm d}N/{\rm dlog}~m$. }
(the number of clusters per {\it logarithmic} mass interval, which is 
proportional to the number of objects per magnitude unit)
is bell-shaped and usually fitted with a gaussian (e.g., Ashman \& Zepf 1998).
However, the underlying mass spectrum (i.e., the number of objects per 
{\it linear} mass interval) is well described by a two-index power-law, 
with exponents $\sim -2$ and  $\sim -0.2$ above and below 
$\sim 1.5 \times 10^5$ M$_{\odot}$, respectively (McLaughlin 1994).   
The peak of the gaussian function in fact coincides with the
cluster mass at which the slope of the mass spectrum changes.
The slope of the high mass regime is reminiscent of what is observed 
in interacting and merging galaxies where systems of young massive
clusters show well defined power-law luminosity spectra 
(i.e., ${\rm d}N \propto L^{-\alpha} {\rm d}L$)
with $\alpha$ in the range 1.8 to 2 (see, e.g., Whitmore \& Schweizer
1995, Whitmore et al.~2002).  This observational fact gave rise to
the popular idea that the initial mass spectrum of the halo 
clusters may have been a power-law with a similar slope.

However, the luminosity spectrum ${\rm d} N/{\rm d} L$ constitutes a
faithfull mirror of the underlying mass spectrum ${\rm d} N/{\rm d} m$   
only if any variations of the mass-to-light ratio from 
cluster to cluster are small.  This is true for any cluster system 
whose stellar initial mass function is invariant and whose cluster
age range is small.  This is roughly the case for the 
Galactic halo globular clusters.  Their visual mass-to-light ratio 
ranges from $\sim$1 to $\sim$4 (see Pryor \& Meylan 1993 and 
Parmentier \& Gilmore 2001, their Fig.~1), partly reflecting
different cluster dynamical evolutions.
This may not be true for cluster systems formed in ongoing or recent 
starbursts.  Their formation duration may be a significant fraction 
of the system's median age and, thus, age spread effects among 
the young star cluster population may not be negligible. 
Being an age related quantity, the cluster integrated mass-to-light ratio
can no longer be considered as a constant even for an invariant
stellar initial mass function.  For instance, a system of young
clusters with an  
age range of 3 to 300 million years shows variations in the cluster
mass-to-light ratio as large as a factor of 20 (see, e.g., Bruzual \&
Charlot 2003).  As a result, the shape of the luminosity 
spectrum may differ substantially from the shape of the mass spectrum 
(Meurer et al.~1995, Fritze v. Alvensleben 1998, 1999).
To unveil the mass spectrum of young clusters therefore requires estimates
of their {\it individual} mass-to-light ratio with, e.g.,  spectral
synthesis models.  However, the results have remained unconclusive so
far: the young cluster system of the nearby 
starburst galaxy NGC~3310 displays a gaussian mass function (and,
thus, a two-index power-law mass spectrum; de Grijs et al.~2003) 
while the young massive clusters located in the Antenna merger 
(NGC~4038/39) are 
distributed in mass according to a pure power-law with slope $\sim -2$ 
(Zhang \& Fall 1999).  These contrasting results may be interpreted as 
evidence against the universality of the globular cluster initial 
mass function.

Whether the Galactic halo cluster system actually started with a
power-law mass  
spectrum or with a gaussian mass function similar to the current one 
has remained a very puzzling issue since both evolve 
into the presently observed lognormal mass function.  As for the
initial power-law mass spectrum, this form gets severely depleted below 
a turnover of $\sim 1.5 \times 10^5$M$_{\odot}$ due to the
preferential removal of low-mass clusters through evaporation
and tidal disruption, leading after a Hubble-time of evolution to the 
current two-index power-law mass spectrum (e.g., Okazaki \& Tosa 1995,
Baumgardt 1998).  On the other hand, Vesperini (1998) demonstrated
that the presently observed gaussian mass function represents a state of
quasi-equilibrium, that is, the gaussian shape and its associated 
parameters (mean and standard deviation) are preserved during the 
entire evolution through a subtle balance between disruption of
clusters and evolution of the masses of those which survive, even
though a significant fraction of the clusters is destroyed.  The
globular cluster initial mass distribution may thus have been a 
gaussian mass function or it may have been a power-law mass spectrum.  
As a result, the low-mass regime cannot be recovered by studying 
the temporal evolution of the mass spectrum only. 
 
Along with the issue of the mass spectrum, several studies dedicated 
to the dynamical evolution of the Galactic globular cluster system 
also address the evolution with time of the cluster radial {\it number}
density profile $n(D)$ (i.e., the number of clusters per unit volume in
space as a function of Galactocentric distance $D$).  Yet, the
temporal evolution of the radial {\it mass} density profile $\rho(D)$
(i.e., the spatial distribution around the Galactic centre of the halo
cluster system mass) is not discussed (e.g., Baumgardt 1998,
Vesperini 1998).  Considering the case of a power-law mass spectrum
with slope $-2$ and extending down to 500\,M$_{\odot}$,
McLaughlin (1999)  emphasized the relative robustness of mass-related
quantities with respect to number-related quantities (see his equations 4-7).
The fact that the total mass and the mass density profile of a cluster
system are
better indicators of the initial conditions than are the number of
clusters and the number density profile will be further illustrated
in our Section~\ref{sec:evol_mod_prof}.  

In contrast to the robustness of the mass density profile,
the disparity in the results derived by Vesperini (1998) and 
Baumgardt (1998) regarding the temporal evolution of the number 
density profile is more puzzling.  The presently observed spatial
distribution of the Galactic halo clusters is centrally concentrated 
with the density varying as $D^{-3.5}$ (Zinn 1985), except in the 
inner 3-4\,kpc where the distribution flattens to something closer to
an $D^{-2}$ dependence.  As this flattening is likely to result (at
least partly) from the shorter time-scale for cluster disruption at smaller
galactocentric distance, Vesperini (1998) assumes that the initial 
number density profile scales as $D^{-3.5}$ through the whole 
halo extent.  Evolving such a system up to an age of 15\,Gyr,  
he concludes that the initial steepness of the distribution is
preserved, except in the inner Galactic regions where the greater
efficiency of cluster destruction processes flattens the profile, in
agreement with the presently observed one.
On the other hand, Baumgardt 's (1998) results suggest that an 
initial slope of $-3.5$ (i.e., a steepness similar to what it is now)
is ruled out as this one leads to a final spatial
distribution  significantly flatter than what is observed.  
Accordingly, the initial distribution must have been steeper and an
initial slope of $\sim -4.5$ is required to match the present spatial
distribution.  It is worth pointing out however that, while Baumgardt
(1998)  builds on a power-law mass spectrum with a slope $\alpha = -2$
and probing down to 1000\,M$_{\odot}$ (i.e., a choice inspired by the 
luminosity spectrum of young massive clusters in starbursts and
mergers), Vesperini (1998) investigates
the case of the equilibrium gaussian mass function (i.e., an initial 
mass function similar to the present one because its shape remains
well-preserved).  Thus, their divergence about the initial steepness
of the number density profile is likely to arise from a different
choice for the initial cluster mass function. 

This also suggests that the mass and number density
profiles evolve differently with time.  While the
former remains a reliable mirror of what it initially was, irrespective
of the initial distribution in mass of the clusters (as we will
confirm in Section \ref{sec:evol_mod_prof}), the steepness of the
latter after a Hubble-time of evolution may depend sensitively on the
initial mass spectrum.  This leads us to consider the possibility 
that a comparison between evolved and observed profiles, in terms of
mass and in terms of number, may help shed light on the initial
mass range and/or the initial distribution in mass of the Galactic
halo clusters. Actually, if we {\it assume} that, soon after their 
formation, globular clusters show the same mass range and the same 
mass spectrum, irrespective of their galactocentric distance, then 
the mass and number density profiles are initially identical in shape.  
On the other hand, the presently observed mass and number density 
profiles of the halo cluster system 
also show shapes that match each other (see, e.g., McLaughlin 1999).
Therefore, the robustness of the mass density profile combined with
(1) the uniformity in shape for the {\it presently observed} mass and
number density profiles and (2) the assumed uniformity in shape for
the {\it initial} ones implies that the initial mass spectrum of
globular clusters was
such that the number density profile has been preserved in spite of
evolution in the radially-dependent tidal fields of the Milky Way.     
   
\begin{figure*}
\begin{minipage}[b]{0.49\linewidth}
\centering\epsfig{figure=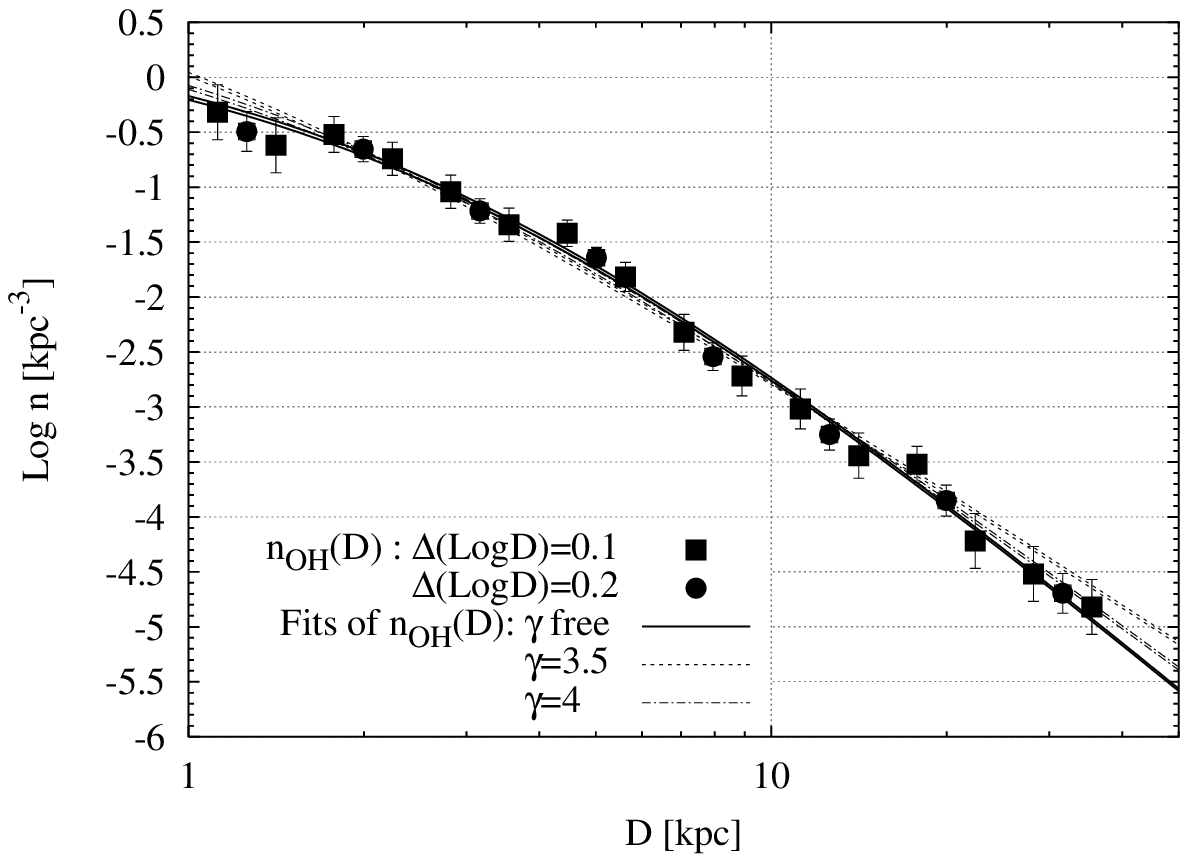, width=\linewidth} 
\end{minipage}
\hfill
\begin{minipage}[b]{0.49\linewidth}
\centering\epsfig{figure=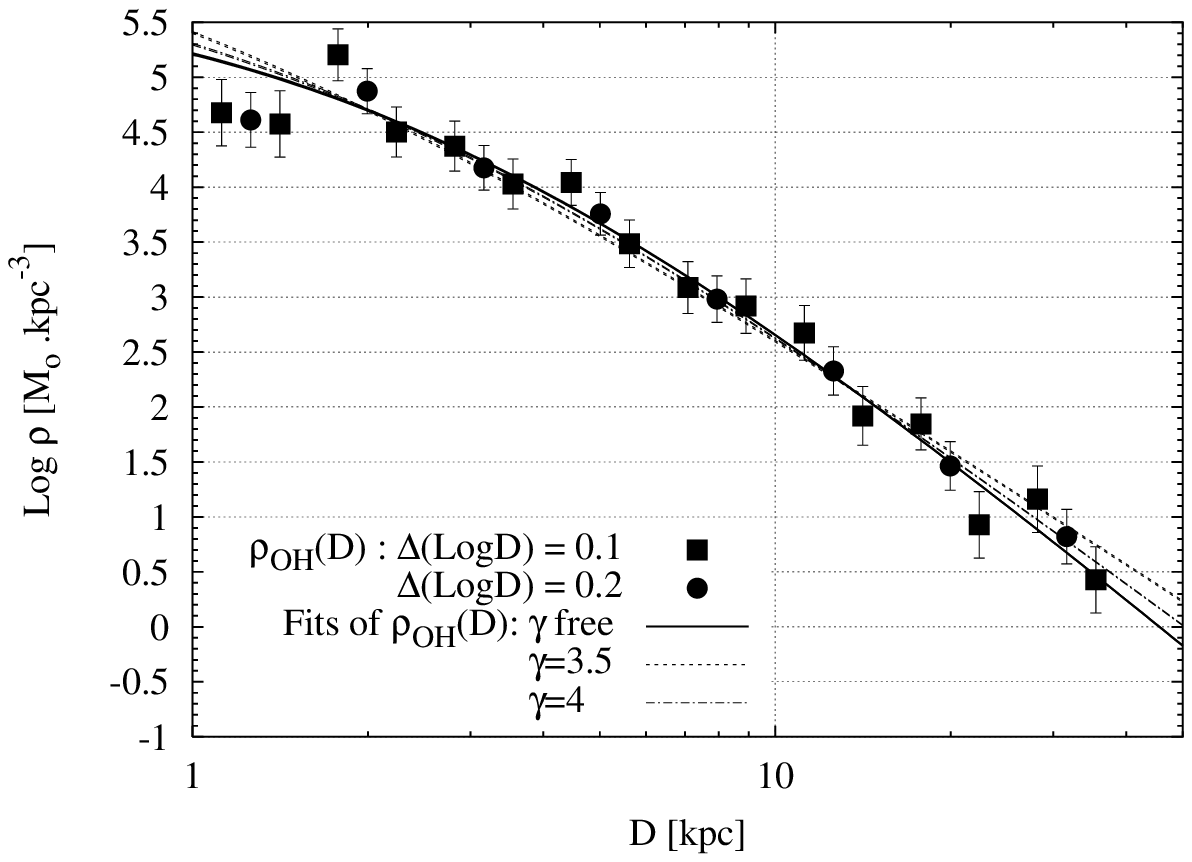, width=\linewidth}
\end{minipage}
\caption{Left panel: fits of a power-law with a core (equation
\ref{eq:log_pl_core}) to the number density profile of the Old Halo
globular cluster subsystem.  The different curves correspond to the
parameter values 
given in Table \ref{tab:fit_pl_core}. Right panel: the same curves
(i.e., same core and same slope) are vertically shifted in order to
match the Old Halo mass density profile.  The good fit to the data in
the right panel shows that the number and the mass density profiles of
the Old Halo are indistinguishable in shape}
\label{fig:prop_er} 
\end{figure*}
   
The outline of the paper is as follows.  In Section 2, we build the 
radial mass and number density profiles of the Old Halo cluster system
and we compare their shape.   
In Section 3, we briefly describe the analytic model as obtained by
Vesperini \& Heggie (1997) which enables us to evolve the number and
mass density profiles of a cluster system.  We evolve various globular
cluster systems with different 
initial mass functions and different initial spatial distributions.
In Section 4, we then compare the model outcomes with the observations 
in order to derive new constraints on the initial mass function of the
Old Halo clusters, as well as on their initial spatial distribution.  
Finally, we present our conclusions in Section 5.     

\section{Radial Profiles of the Old Halo}
\label{sec:OH_prof}

Comparing the slopes of the mass density profile and of the number
density profile, which is 
the core of this paper, is equivalent to studying the radial variation
of the mean cluster mass (and, in fact, gives the physical reason for
such a variation, namely, the possibly different responses to the
dynamical evolution of the number-related and mass-related quantities).  
The observed radial variations of the mean
cluster mass in the Milky Way, M31 and M87 have
already been tackled by Gnedin (1997).  Sorting the clusters
in two equal size parts with respect to their galactocentric distance,
he detected the existence of ``statistically significant differences
between the inner and outer populations, inner clusters being on
average brighter than the outer clusters, as would be expected if the
inner population had been depleted by tidal shocks.'' 
The dynamical interpretation of the results is presented in a
complementary study by Ostriker \& Gnedin (1997).  However, his
Milky Way sample does not discriminate among the different cluster
populations, that is, disc/bulge, Old Halo and Younger Halo (see
below).  Due to obvious differences with respect to their age,
their metallicity and their evolution history (i.e., accreted clusters
have not been constantly subjected to the Galactic potential since the
time of their formation), the interpretation of differences in the
mean luminosity of the inner and outer populations may not be that
straightforward.  As for M87, the conclusions of Gnedin (1997) and
Ostriker \& Gnedin (1997) have been revisited by Kundu et al.~(1999)
who detected no significant variations of the cluster luminosity
function turnover with respect to the projected galactocentric
distance.   Kundu et al.~(1999) claimed that ``the apparent
brightening observed by Gnedin is probably due to undercompensation of 
completeness corrections in the inner regions, where the dimmer
clusters are harder to detect against the strong galaxy background.''     
In contrast to Kundu et al.'s (1999) results, Barmby et
al.~(2001) found that the mean luminosity of the inner clusters of M31
is significantly brighter than that of the outer ones.  However, they
cautionned that variations driven by differences in cluster
metallicity, age and stellar initial mass function may also be
important and must be accounted for properly before using any
luminosity function variation as a probe to differences in the globular
cluster mass function. 

\begin{table*}
\begin{center}
\caption[]{Results of fitting a power-law with a slope $-\gamma$ and
a core $D_c$ (equation \ref{eq:log_pl_core}) on the observed number 
density profile of the Old Halo subsystem.  The core and exponent derived
for the number density profile are then directly applied to the mass
density profile, the corresponding $\chi ^2$ and $Q(\nu /2, \chi
^2/2)$ being provided in the last two columns}  \label{tab:fit_pl_core}
\begin{tabular}{ l c c c r c c c r c} \hline 
$\Delta {\rm Log} D$~~ & $Log(n _0)$ & $D_c$  & $-\gamma$ &
$\chi ^2$~ &  Q & ~~~ & $Log(\rho _0)$ & $\chi ^2$~ &  Q \\ \hline
0.1   & 0.53 $\pm$ 0.32 &  2.33 $\pm$ 1.03 & --4.50 $\pm$ 0.53 & 6.63
& 0.92 & & 5.92 $\pm$ 0.06 & 12.68 & 0.63 \\ 
0.2   & 0.49 $\pm$ 0.33 &  2.33 $\pm$ 1.08 & --4.50 $\pm$ 0.55 & 4.53
& 0.48 & & 5.90 $\pm$ 0.08 & 4.74 & 0.69 \\ \hline 
      &              &               &$\gamma$ imposed &     &     \\ \hline  
0.1   & 1.42 $\pm$ 0.41 &  0.68 $\pm$ 0.21 & --3.5          &
12.63 & 0.56 & & 6.79 $\pm$ 0.06 & 18.28 & 0.25 \\ 
0.2   & 1.38 $\pm$ 0.42 &  0.69 $\pm$ 0.22 & --3.5          &
10.06 & 0.12 & & 6.77 $\pm$ 0.08 & 7.87 & 0.34 \\ \hline
0.1   & 0.84 $\pm$ 0.24 &  1.45 $\pm$ 0.26 & --4.0          &
7.84 & 0.90 & & 6.22 $\pm$ 0.06 & 14.44 & 0.49 \\ 
0.2   & 0.80 $\pm$ 0.25&   1.46 $\pm$ 0.27 & --4.0
&5.63 & 0.47 & & 6.20 $\pm$ 0.08 & 5.64 & 0.58 \\ \hline 
\end{tabular}
\end{center}
\end{table*}   

\begin{table*}
\begin{center}
\caption[]{Results of fitting a power-law $D^{-\gamma}$ to the number
and mass density profiles of the Old Halo subsystem, considering
clusters with $D \gtrsim 3$\,kpc only, that is, focusing on regions 
which have been least affected by dynamical evolution} 
\label{tab:fit_pure_pl}
\begin{tabular}{ l c c c r c c c r c} \hline 
$\Delta {\rm Log} D$~~ & $Log(n _0)$ & $-\gamma$ & $\chi ^2$~ &  Q &
~~~ & $Log(\rho _0)$ & $-\gamma$ & $\chi ^2$~ &  Q \\ \hline
0.1   & 0.83 $\pm$ 0.17 &  --3.64 $\pm$ 0.18 & 5.00
& 0.83 & & 6.26 $\pm$ 0.25 & --3.66 $\pm$ 0.24 & 6.11 & 0.73 \\ 
0.2   & 0.68 $\pm$ 0.15 & --3.53 $\pm$ 0.16 & 4.66 & 0.32 & & 6.06 $\pm$
0.27 & --3.46 $\pm$ 0.26 & 1.33 & 0.86 \\ \hline 
\end{tabular}
\end{center}
\end{table*}   

In the present study, we restrict our attention to the Galactic 
globular cluster system.  In the Galaxy 
only can we clearly discriminate among the different cluster
populations regarding their age and evolutionary history.  
Specifically, we aim at constraining the initial mass spectrum of the
first generation globular clusters which formed within the 
gravitational potential well of the Galaxy.  Hence, we do not consider 
the more metal-rich, presumably second-generation, bulge/disc 
globular clusters (Zinn 1985).
The halo cluster system itself hosts two distinct populations of clusters,
the so-called Old Halo and Younger Halo (Van den Bergh 1993,
Zinn 1993).  The Old Halo globular clusters might have been formed 'in
situ'.  In contrast, younger halo globular clusters
are suspected of having been accreted.  Regardless of their
formation history, the Old Halo globular clusters form a coherent
and well-defined group, well-suited to an analysis of their
properties: thus, we consider the Old Halo subgroup only.
Lists were initially compiled by Lee et al.~(1994) and 
Da Costa \& Armandroff (1995) and have recently been updated by Mackey 
\& Gilmore (2004).  We note in passing that, although coeval with the 
inner halo (Harris et al., 1997), we have not included NGC~2419
in our Old Halo sample.  This cluster is located at a 
galactocentric distance of order 90\,kpc and is thus unlikely to belong
to the main body of the Galaxy.  Moreover, van den Bergh \& Mackey (2004)
show that NGC~2419 and $\omega$ Cen on the one hand, and the other
halo clusters
on the other hand, are at different locii in a half-light radius vs 
absolute visual magnitude diagram.  This thus suggests that, 
as for $\omega$ Cen (which we also exclude from our sample), NGC~2419
might be the tidally stripped core of a former dwarf spheroidal galaxy.

To build the mass and number density profiles, our source for the 
galactocentric distances $D$ and the absolute visual magnitudes 
M$_v$ is the McMaster database compiled and maintained 
by Harris (1996, updated February 2003). 
Cluster absolute visual magnitudes have been turned into luminous
mass estimates by assuming a constant mass-to-light ratio M/L$_v$=2.35
(i.e., the average of the mass-to-light ratios of the
halo clusters for which Pryor \& Meylan (1993) obtained dynamical mass
estimates).  
The Old Halo mass and number profiles are derived by binning the data with
two different  
bin sizes: $\Delta {\rm log} D = 0.1~ {\rm and}~ 0.2$ ($D$ is in kpc), 
corresponding to 16 and 8 points, respectively.
As for the size of the error bars, a Poissonian error on the number of 
clusters in each bin is combined with a fixed error on the
mass-to-light ratio.  In fact, not all globular clusters show the same 
mass-to-light ratio, the standard deviation in the Pryor \& Meylan
(1993) compilation being of order $\sigma _{{\rm Log}(M/L_v)} = 0.17$.  \\  

As already mentioned, the observed radial distribution of halo clusters
obeys $D^{-3.5}$ except in the innner 3-4\,kpc where the distribution
gets shallower.  As a result, it is often parametrized by a power law
with a core (see equation \ref{eq:log_pl_core}).   Previous fits
having been obtained for either the whole Galactic globular cluster
system (e.g., 
Djorgovski \& Meylan 1994) or the whole halo system (e.g., McLaughlin
1999), we now consider the Old Halo subsystem only.  Using a
Levenberg-Marquardt algorithm (Press et al.~1992), we fit the 
Old Halo number density profile with:
\begin{equation}
{\rm Log} n(D) = {\rm Log} n_0 - \gamma \, {\rm Log} \left(1+\frac{D}{D_c} \right)\;.
\label{eq:log_pl_core}
\end{equation}

The values obtained for the slope $-\gamma$ and the core $D_c$ are
presented in the left part of Table \ref{tab:fit_pl_core}.  For each
fit, we also give the $\chi ^2$ and
the incomplete gamma function $Q(\nu /2, \chi ^2/2)$ ($\nu$ is the 
number of degrees of freedom) which provides a quantitative measure
for the goodness-of-fit of the model \footnote{We remind the reader
that: a $Q$ value of 0.1 or larger indicates a satisfactory agreement
between the model and the data; if $Q \geq 0.001$, the fit may be
acceptable if, e.g., the errors have been moderately underestimated;
if $Q < 0.001$, the model can be called into question (Press\ et
al.~1992).}.  
Imposing a slope $-\gamma$ of --3.5 or --4, as found by Djorgovski \&
Meylan (1994) for the whole globular cluster system, provides a good
fit to the number 
density profile of the Old Halo cluster system as well.  Keeping all three
parameters free, we obtain a steeper slope ($-\gamma \simeq -4.5$)
coupled with a larger core.  The core reflects the flattening of the
spatial distribution at small galactocentric distances, presumably
owing to the 
greater efficiency of disruptive processes.  Ignoring this core region and
focusing on the Old Halo clusters located at galactocentric distances $\gtrsim
3$\,kpc, that is, where memory of the initial conditions has perhaps
been better preserved, we find that both the mass and the number 
density profiles of the Old Halo are well-approximated by pure power-laws
with slope $\simeq -3.5$ (see Table \ref{tab:fit_pure_pl}).  
The steepness of the Old Halo
spatial distribution is thus similar to that of the whole halo
(Zinn 1985).  While the mass and number density profiles show very similar
steepness at distances larger than 3\,kpc, their overall shapes are
also very similar.  Fitting the Old Halo mass density profile with the same
functions as used for the number density profile (i.e., equation 
\ref{eq:log_pl_core} and the ($\gamma$, $D_c$) couples listed in Table
\ref{tab:fit_pl_core}) provides equally good values of the incomplete 
gamma function (see the last column of Table \ref{tab:fit_pl_core}).  
Therefore, the number and the mass density profiles of the Old Halo
are indistinguishable through the whole extent of the halo. 

\section{Evolved Radial Density Profiles of Globular Cluster Systems}
\label{sec:evol_mod_prof}

The previous section shows that the {\it currently observed} mass
density and number density profiles of the Old Halo are identical in
shape.  If we {\it assume} that the globular cluster formation 
mechanism produced the  same mass range and the same mass spectrum 
for the clusters, irrespective of their galactocentric
distance, then the {\it initial} mass and number density profiles
were identical in shape as well.  If the mass density profile has
been preserved (and we show in this section that it is
actually the case), all together, these facts imply that the number
density profile itself has remained fairly unaltered during
evolution in the tidal field of the Milky Way.  
In what follows, we evolve various putative globular cluster systems, 
considering different combinations of initial mass spectra and initial 
spatial distributions. 
We then investigate in which case(s) has the number density
profile been reasonably preserved.  We also compare
in a least-squares sense the evolved spatial distributions to the
observed ones that we have derived in Section 2. \\

To evolve the radial mass and number density profiles of a cluster
system from the time of its formation up to an age of 15\,Gyr, 
we adopt the analytic formula of Vesperini \& Heggie (1997) which
supplies at any time $t$ the mass $m$ of a star cluster with initial 
mass $m_i$ which is moving along a circular orbit perpendicular to the
galactic disc at a galactocentric distance $D$.  
The assumption of circular orbits is clearly a simplifying one since
it implies that the time variations of the tidal field for clusters on
elliptical orbits are not allowed for in our calculations.  Yet, the
system of relevance here is the Old Halo.  This shows less extreme
kinematics than the Younger Halo group of clusters, making this
assumption less critical than if we have dealt with the whole
halo system.  As for the influence of the cluster orbit inclination with
respect to the Galactic disc, Murali \& Weinberg (1997) found that, although
low-inclination halo clusters evolve more rapidly than
high-inclination ones, the differences are not extreme.
Furthermore, our sample excluding disc clusters, the 
assumption of high inclination orbits is a reasonable one. 
The simulations of Vesperini \& Heggie (1997) were designed
in the frame of a host galaxy modelled as a simple isothermal sphere
with a constant circular velocity.  This actually constitutes a reasonable
assumption for the Old Halo system whose radial extent is 1-40\,kpc,
that is, where the mass profile of the Milky 
Way (i.e, the total Galactic mass enclosed within a radius $D$) grows
linearly with the galactocentric distance (Harris 2001).  We consider
the effect of non-circular orbits in more detail in Section 4,
below.
  
The relations describing the temporal evolution of the mass of a
globular cluster have been obtained by fitting 
the results of a large set of N-body simulations in which Vesperini \& 
Heggie (1997) take into account the effects of stellar evolution as well as
two-body relaxation, which leads to evaporation through the cluster tidal 
boundary.  Disc shocking can also be included 
(see below).  In order to take into account dynamical friction,
globular clusters whose time-scale of orbital decay (see, e.g., 
Binney \& Tremaine 1987) is smaller than $t$ are removed from the 
cluster system at that time (see Vesperini 1998, his Section 2, 
for further details). 
It is important to note a specific assumption underlying the validity
of this analysis.  The large-scale Galactic gravitational potential is
assumed constant, that is, this model considers the evolution of a
globular cluster system only after it has been assembled in 
a time-independent Galaxy.  That is the physical basis for a restriction
to the system of Old Halo globular clusters.  

The temporal evolution of the mass of a cluster
orbiting at constant galactocentric distance $D$ is found to follow:
\begin{equation}
\frac{m(t)}{m_i} = 1 - \frac{\Delta m_{st,ev}}{m_i}
- \frac{0.828}{F_{cw}} t\;.
\label{eq:mgc_t_noDS}
\end{equation} 

$\Delta m_{st,ev}/m_i$ is the fraction of cluster
mass lost due to stellar evolution (18 per cent in this particular model).
The time $t$ is expressed in units of 1\,Myr and $F_{cw}$, a quantity 
proportional to the initial relaxation time, is defined as:
\begin{equation}
F_{cw} = \frac{m_i \times D}{{\rm ln} N}\,,
\label{eq:Fcw}
\end{equation}

where $m_i$ and $D$ are in units of 1\,M$_{\odot}$ and 1\,kpc,
respectively, and $N$ is the initial number of stars in the cluster.
To take into account disc shocking, the factor 0.828/$F_{cw}$ is merely 
replaced by $\lambda$ as defined by equation (3) of Vesperini (1998). \\  

\begin{table}
\begin{center}
\caption[]{Fraction F$_N$ of surviving clusters and ratio F$_M$ of the
final to the initial mass in clusters after a 15\,Gyr long evolution for
globular cluster systems with various initial spatial distributions 
$n _{init}$(D) and cluster initial mass functions/spectra (CIMF/Sp).  
G refers to a gaussian mass function ${\rm d}N/{\rm 
d~log}~m$ with a mean ${\rm log}m_0=5.03$ and a standard deviation
$\sigma =0.66$ (i.e., the equilibrium mass function as defined by
Vesperini 1998).   PL5, PL4 and PL3 refer to power-law mass spectra
${\rm d}N/{\rm d}m$, with a slope $\alpha =-1.9$ and extending down to
1E5\,M$_{\odot}$, 1E4\,M$_{\odot}$ and 1E3\,M$_{\odot}$, respectively.
Results are presented with (DS) and without (no DS) disc shocking
included in the simulations}  \label{tab:frac_surv}
\begin{tabular}{ l c c c c c c } \hline 
 & & \multicolumn{2}{c}{no DS} & & \multicolumn{2}{c}{~~~DS} \\ \hline 
$n _{init}$(D) & CIMF/Sp & F$_M$ & F$_N$ &  & F$_M$ & F$_N$  \\ \hline 
D$^{-3.5}$        & G   & 0.54 & 0.62 & & 0.50 & 0.56 \\ 
                  & PL5 & 0.55 & 0.87 & & 0.51 & 0.77 \\ 
                  & PL4 & 0.39 & 0.32 & & 0.36 & 0.30 \\
                  & PL3 & 0.30 & 0.05 & & 0.28 & 0.04 \\ \hline 
D$^{-4.5}$        & G   & 0.37 & 0.43 & & 0.31 & 0.35 \\ 
                  & PL5 & 0.38 & 0.72 & & 0.32 & 0.55 \\ 
                  & PL4 & 0.24 & 0.14 & & 0.20 & 0.11 \\
                  & PL3 & 0.19 & 0.02 & & 0.15 & 0.01 \\ \hline 
\end{tabular} 
\end{center}
\end{table}   

We have distributed 20,000 clusters following various radial
and mass distributions.  Four different initial distributions in
cluster mass have been considered:
\begin{itemize}
\item[a) ] a gaussian mass function ${\rm d}N/{\rm dlog}m$ with parameters 
equal to those of the equilibrium mass function of Vesperini (1998),
that is, a mean ${\rm log}m_0=5.03$ and a standard deviation
$\sigma =0.66$;
\item[b) ] three power-law mass spectra ${\rm d}N/{\rm d}m$, each with a 
slope of --1.9 and different lower mass limits, namely, 1E3, 1E4 
and 1E5\,M$_{\odot}$.  The value of the slope agrees with what is
obtained for the high mass regime of the Galactic halo cluster system, 
that is, a slope of around --1.8 to --2 (see, e.g., Ashman \& Zepf 1998).     
\end{itemize}
As for the last low-mass cut-off, Fall \& Zhang (2001) indeed show
that the cluster mass spectrum may have started with a truncation at
mass of order 1E5\,M$_{\odot}$, the low-mass tail of the present mass
distribution resulting from the evaporation of initially more massive
clusters located close to the
Galactic centre.  This illustrates one more time the difficulty of
deducing the initial distribution in mass of the globular clusters on 
the sole ground of evolving it with time.  Three different initial mass
spectra manage to evolve into the presently observed bell-shaped mass
function: (1)~a power-law probing down to 1000\,M$_{\odot}$ or
(2)~truncated at $10^5$\,M$_{\odot}$ as well as (3)~a two-index
power-law with a turnover around $10^5$\,M$_{\odot}$.  We thus are 
ignorant of the contribution of the low-mass objects to the initial
population of globular clusters.  Did they constitute the overwhelming 
contribution by number (case 1), were they missing (case 2), or were 
they present in limited number (case 3) ?     \\

Regarding the initial number density profile (equivalent in shape to
the initial mass density profile following our assumption of a unique
mass range and a unique mass spectrum through the Old Halo extent), we assume
two different functional forms:
\begin{itemize}
\item[a) ] $n(D) \propto \rho(D) \propto D^{-3.5}$, as suggested by
our fits to the observed spatial distributions for globular clusters 
located beyond 3\,kpc (see Table \ref{tab:fit_pure_pl});
\item[b) ] $n(D) \propto \rho(D) \propto D^{-4.5}$ (Baumgardt 1998).
\end{itemize}
We have thus considered a total of 16 different cases, combining the 4
different initial mass spectra/functions with the 2 different initial
spatial-densities and including or not disc shocking. 

\begin{table*}
\begin{center}
\caption[]{Results of fitting the radial distributions evolved over
15\,Gyr to the mass and number density profiles of the Old Halo
globular cluster system, for the two
different bin sizes considered ($\Delta {\rm Log} D = 0.1$ or 0.2).
The corresponding $\chi ^2$ and $Q(\nu /2, \chi ^2/2)$ (i.e., the
goodness-of-fit) are given for each of the sixteen different cases considered}
\label{tab:fit_evol_GCS}
\begin{tabular}{ l r c c c c c c c c c c c} \hline 
\multicolumn{2}{l}{} &  \multicolumn{5}{c}{$n _{init}$(D) $\propto$ D$^{-3.5}$} & ~~~~~ & \multicolumn{5}{c}{$n _{init}$(D) $\propto$ D$^{-4.5}$}  \\ \hline  
           &                      & \multicolumn{2}{c}{$\rho$(D)} & &
           \multicolumn{2}{c}{$n$(D)}  & & \multicolumn{2}{c}{$\rho$(D)} & & \multicolumn{2}{c}{$n$(D)}  \\ \hline
\multicolumn{2}{c}{$\Delta {\rm Log} D = $} & 0.1   & 0.2  & & 0.1   & 0.2 & & 0.1   & 0.2  & & 0.1   & 0.2  \\ \hline  
G          &                      &       &      & &       &      &    &  &      & &      & \\  
NoDS        & $\chi ^2 =~~~ $            & 13.05 & 4.53 & & 11.84 & 9.40 & & 63.49 & 44.79  & & 87.48 & 82.56     \\  
           & $Q =~~~ $                  & 0.60  & 0.72 & & 0.69  & 0.23 & & $6 \times 10^{-8}$ & $2 \times10^{-7}$ & & $3 \times 10^{-12}$ & $4 \times 10^{-15}$ \\
~~~DS        & $\chi ^2 =~~~ $            & 11.13 & 2.32 & & 9.97  & 7.27 & & 50.79 & 36.11 & & 67.35 & 64.51  \\  
           & $Q =~~~ $                  & 0.74  & 0.94 & & 0.82  & 0.40 & & $9 \times 10^{-6}$ & $7 \times 10^{-6}$ & & $1 \times 10^{-8}$ & $2 \times 10^{-11}$ \\  \hline 
PL5        &                      &       &      & &       & &    & &           & &      & \\  
NoDS          & $\chi ^2 =~~~ $            & 13.19 & 4.64 & & 16.76 &
           14.74 &  & 64.68 & 45.60 & & 127.5 & 122.0  \\  
           & $Q =~~~ $                  &  0.59 & 0.70 & & 0.33 & 0.04   & & $4 \times 10^{-8}$ & $1 \times 10^{-7}$ & & $7 \times 10^{-20}$ & $3 \times 10^{-23}$\\  
~~~DS        & $\chi ^2 =~~~ $            & 10.93 & 2.19 & & 9.45  & 6.90  & &           50.3 & 35.7 & & 99.6 & 94.7 \\  
           & $Q =~~~ $                  & 0.76  & 0.95 & & 0.85 & 0.44   & & $1 \times 10^{-5}$ & $8 \times 10^{-6}$ & & $2 \times 10^{-14}$ & $1 \times 10^{-17}$\\  \hline 
PL4        &                      &       &      & &      & &      & &           & &      & \\  
NoDS       & $\chi ^2 =~~~ $            & 13.71 & 4.39 & & 42.31 & 37.85 & &  51.53 & 35.76  & & 25.37 & 23.18  \\  
           & $Q =~~~ $                  & 0.55  & 0.73 & &    $2 \times 10^{-4}$ & $3 \times 10^{-6}$ & & $7 \times 10^{-6}$ &
           $8\times10^{-6}$ & & 0.05 & $2 \times10^{-3}$ \\  
~~~DS        & $\chi ^2 = ~~~$            & 13.15 & 3.00 & & 57.97  & 51.88  &  & 38.55 & 26.76 & & 13.73 & 12.73  \\  
           & $Q =~~~ $                  & 0.59  & 0.89 & &  $6 \times 10^{-7}$ & $6 \times 10^{-9}$ & & $8 \times 10^{-4}$ & $4
           \times 10^{-4}$ & & 0.55 & 0.08 \\  \hline 
PL3        &                      &       &      & &      & \\  
NoDS       & $\chi ^2 =~~~ $            & 13.90 & 4.45 & & 52.60 & 47.15 & & 51.23 & 35.48  & & 23.63 & 20.90  \\  
           & $Q =~~~ $                  & 0.53  & 0.73 & & $4 \times 10^{-6}$ & $5 \times 10^{-8}$ & & $8 \times 10^{-6}$ & $9\times 10^{-6}$ & & 0.07 & $4 \times 10^{-3}$ \\  
~~~DS        & $\chi ^2 =~~~ $            & 13.37 & 3.09 & & 69.32  & 62.31 & & 38.27 & 26.51  & & 13.11  & 11.54     \\  
           & $Q =~~~ $                  & 0.57  & 0.88 & & $6 \times 10^{-9}$ & $5 \times 10^{-11}$ & & $8 \times 10^{-4}$ & $4 \times 10^{-4}$ & &  0.59 & 0.12 \\  \hline 
\end{tabular}
\end{center}
\end{table*}   

Table \ref{tab:frac_surv} lists the fraction of surviving clusters
($F_N$) and the ratio of the final to the initial mass in clusters 
($F_M$) for each case.  For a given initial space-density, we note
the relative homogeneity of the mass fractions
$F_M$ in spite of widely different initial mass spectra, the extreme
values differing by a factor of 2 at most.  Also, as noted by previous
studies (Vesperini 1998, Baumgardt 1998, McLaughlin 1999), the
evaporation and the disruption of globular clusters cannot account for the
overwhelming contribution of field stars to the luminous
Galactic halo.  The mass of the Old Halo cluster subsystem is $\simeq 2 \times
10^7$\,M$_{\odot}$, that is, about two per cent only of the stellar
halo mass ($\simeq 10^9$\,M$_{\odot}$, Freeman \& Bland-Hawthorn 2002).  
On the other hand, the largest destruction rates in Table
\ref{tab:frac_surv} (F$_N =$ 0.01-0.02) correspond to a total mass in
survivors of order 15 to 20 per cent of the initial cluster system
mass.  Hence, even in this extreme case, disrupted and evaporated clusters
account for 10 to 13 per cent of the stellar halo mass only. 

In contrast to the rather limited dispersion shown by the mass
fraction, $F_M$, the fraction of survivors in the number density
distribution $F_N$ is characterized by a scatter as large as an
order of magnitude for both initial spatial distributions.  
Larger destruction rates are of course achieved in case of a
power-law initial mass spectrum probing down to 1000\,M$_{\odot}$
as this one favours low-mass easily disrupted clusters.  
The fraction of survivors is also smaller in case of a
steeper initial spatial distribution (i.e., D$^{-4.5}$ instead of
D$^{-3.5}$) since a larger fraction of globulars are then located 
at smaller galactocentric distances where destruction processes 
proceed on a shorter time-scale.  

\begin{figure*}
\begin{minipage}[b]{0.48\linewidth}
\epsfig{figure=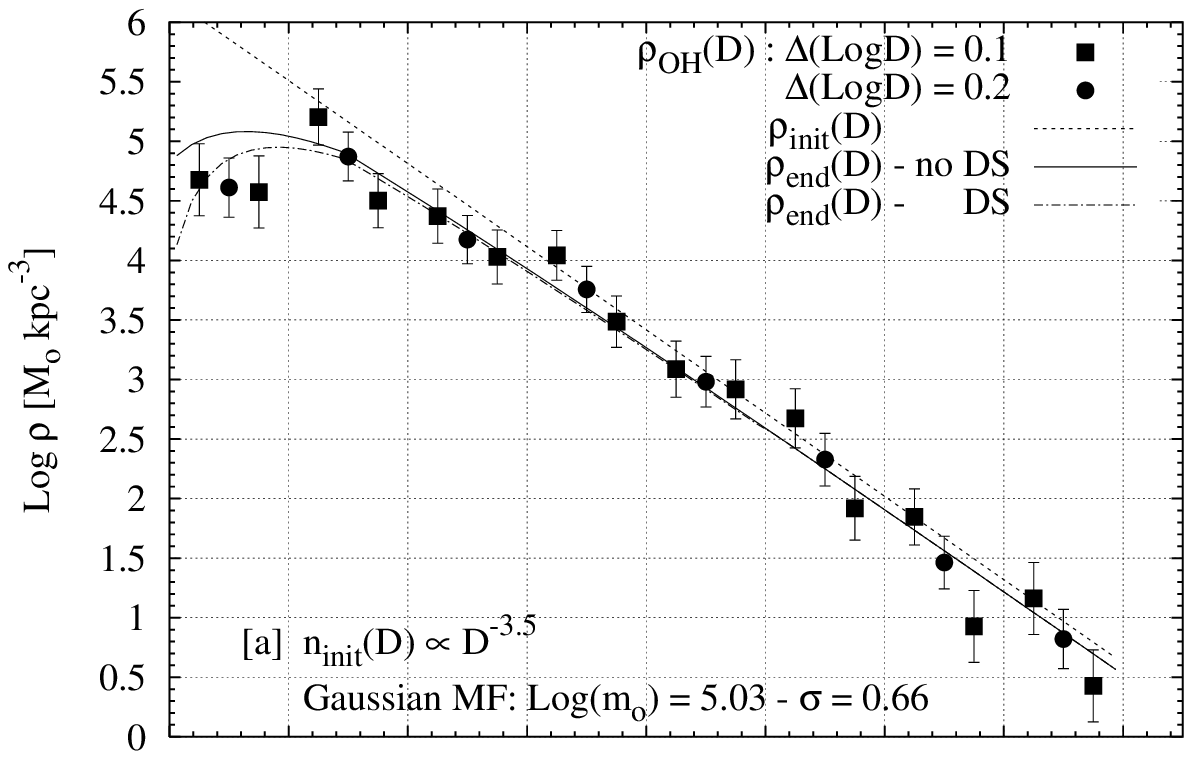, width=\linewidth}
\end{minipage}
\hfill
\begin{minipage}[b]{0.48\linewidth}
\epsfig{figure=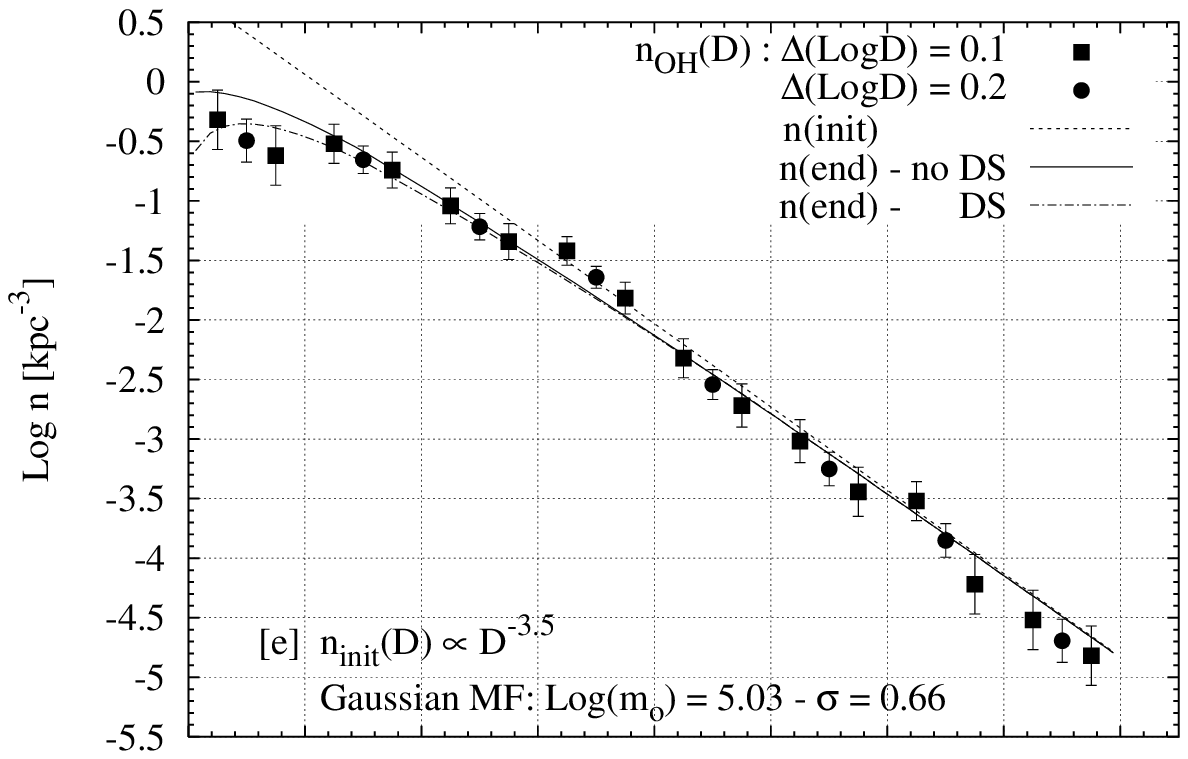, width=\linewidth}
\end{minipage}
\vfill
\vspace*{-8mm}
\begin{minipage}[b]{0.48\linewidth}
\epsfig{figure=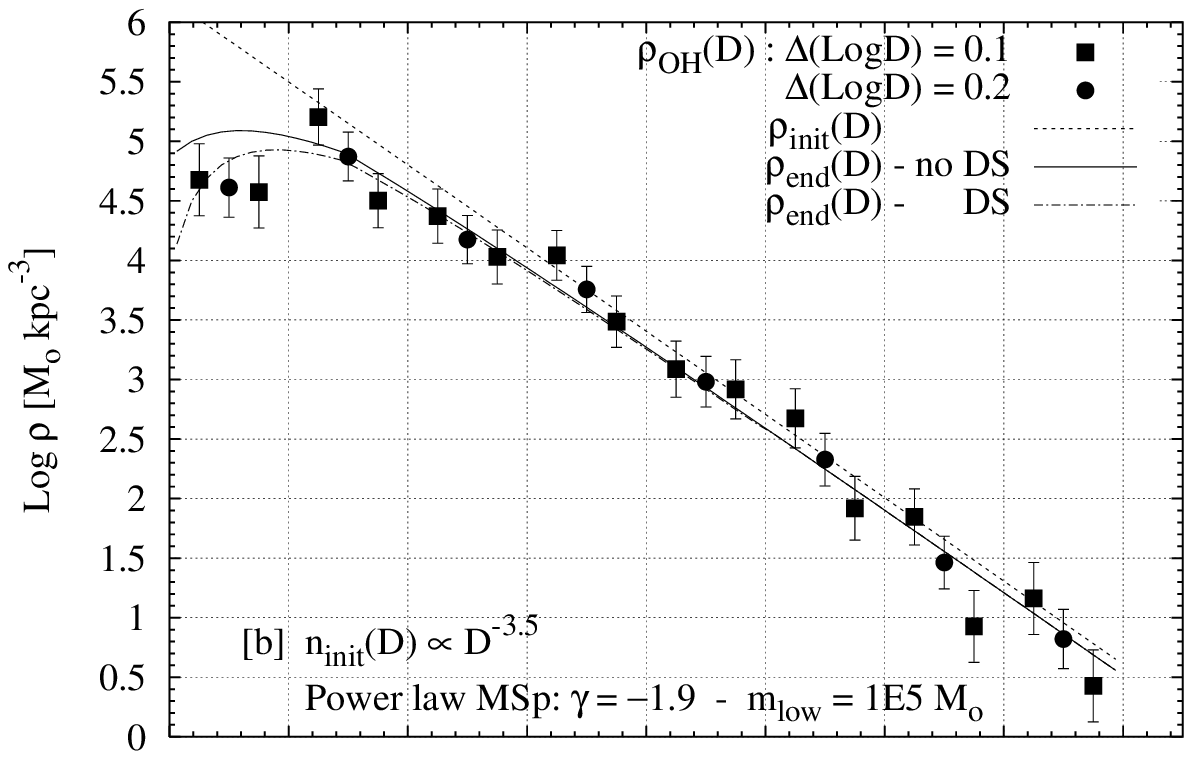, width=\linewidth}
\end{minipage}
\hfill
\begin{minipage}[b]{0.48\linewidth}
\epsfig{figure=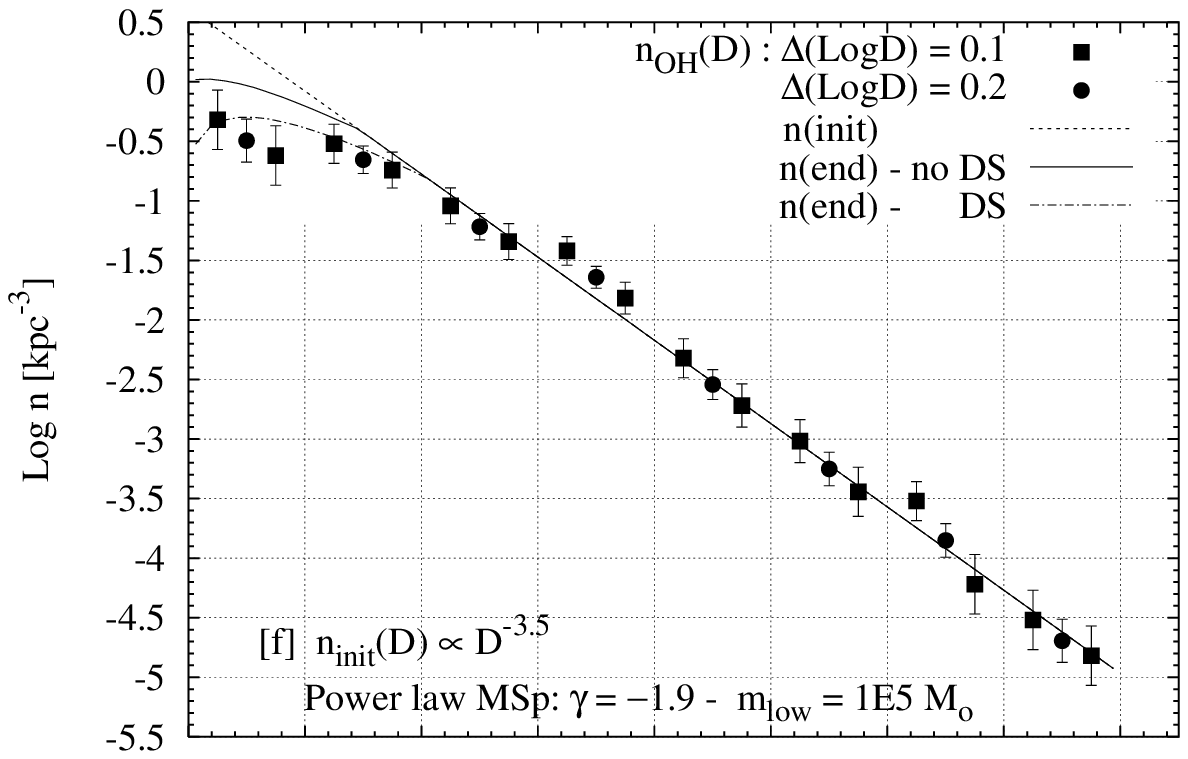, width=\linewidth}
\end{minipage}
\vfill
\vspace*{-8mm}
\begin{minipage}[b]{0.48\linewidth}
\epsfig{figure=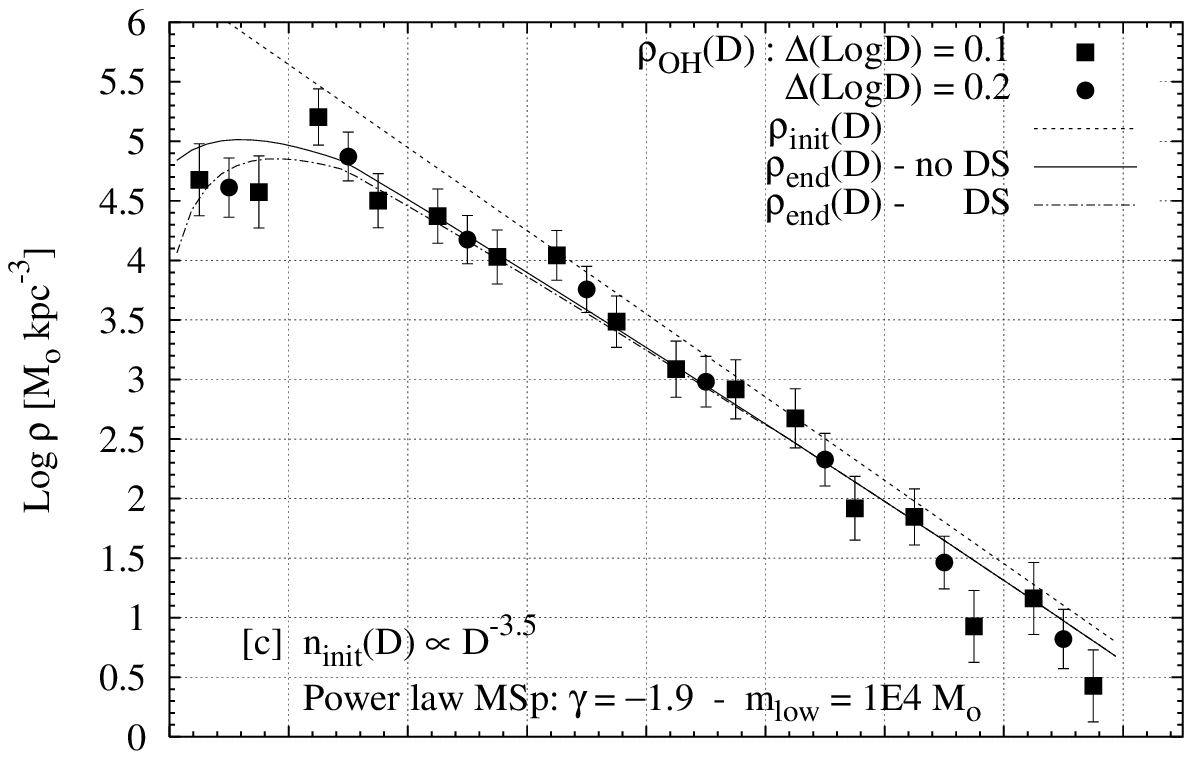, width=\linewidth}
\end{minipage}
\hfill
\begin{minipage}[b]{0.48\linewidth}
\epsfig{figure=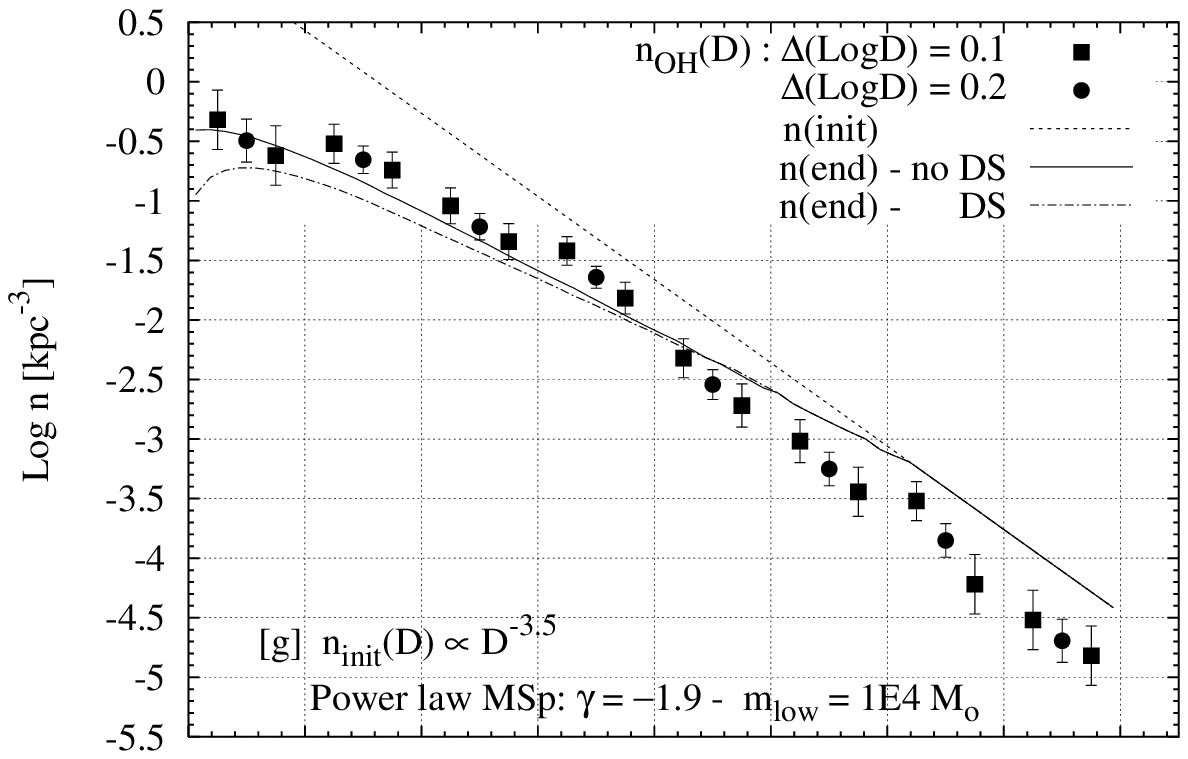, width=\linewidth}
\end{minipage}
\vfill
\vspace*{-8mm}
\begin{minipage}[b]{0.48\linewidth}
\epsfig{figure=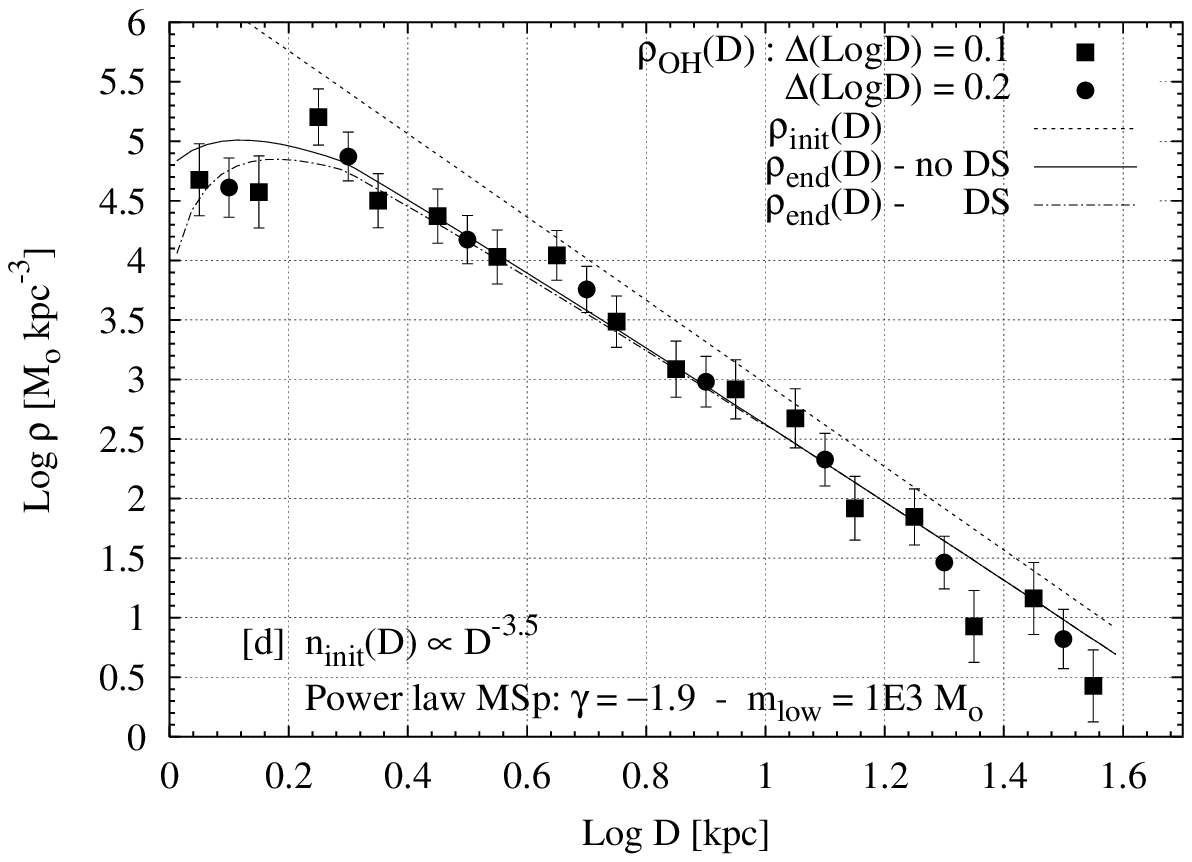, width=\linewidth}
\end{minipage}
\hfill \hspace*{5mm}
\begin{minipage}[b]{0.48\linewidth}
\epsfig{figure=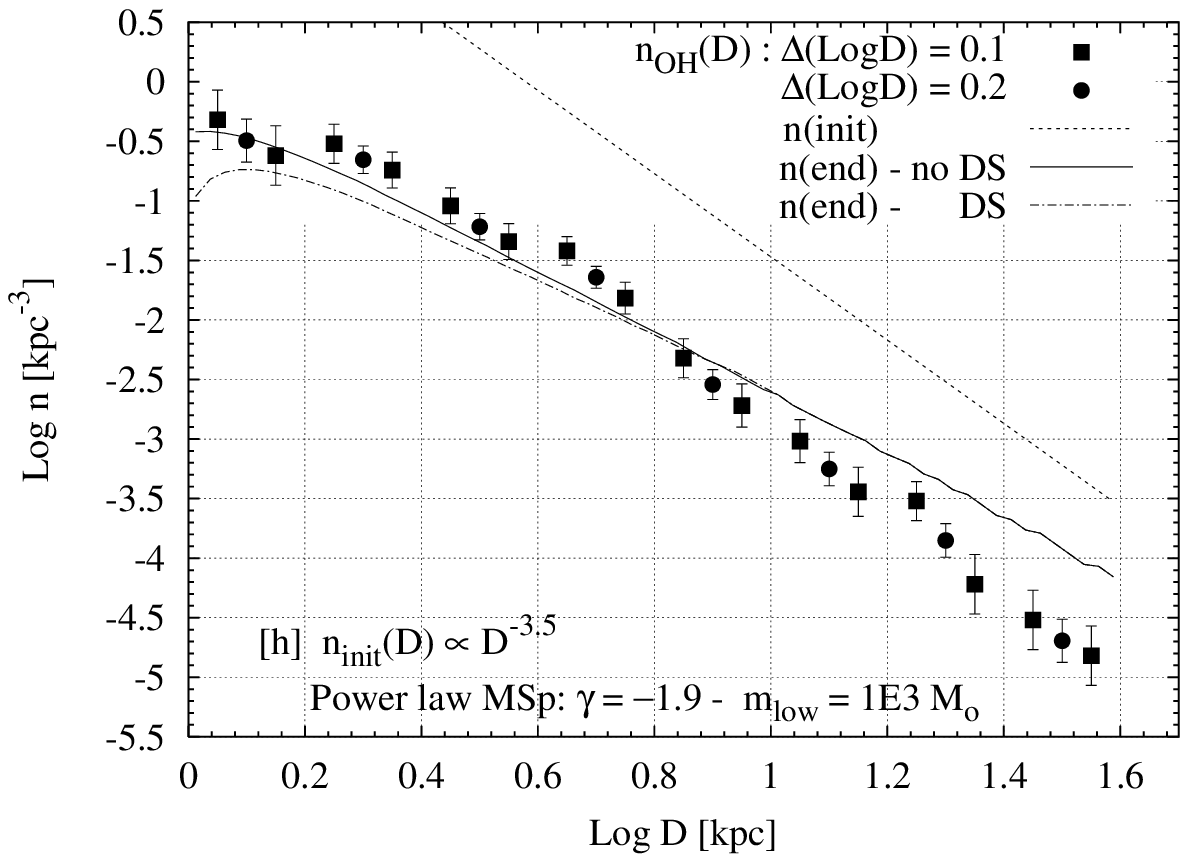, width=\linewidth}
\end{minipage}
\caption{Evolution with time of the radial mass (left panels) and
number (right panels) density profiles of globular cluster systems 
whose initial space-density (dotted curves) obeys $n_{init} \propto D^{-3.5}$.
Different initial distributions in mass are used, from top to bottom: a
gaussian mass function ${\rm d}N/{\rm d~log}~m$ with a mean ${\rm
log}m_0=5.03$ and a standard deviation $\sigma =0.66$, 3 power-law
mass spectra ${\rm d}N/{\rm d}m$ with a slope $\alpha 
=-1.9$ and extending down to 1E5\,M$_{\odot}$, 1E4\,M$_{\odot}$ and
1E3\,M$_{\odot}$, respectively. Results with (dashed-dotted curves) and
without (solid curves) disc-shocking are shown.  The solid curves are
vertically shifted in order to provide the best match to the Old Halo data}  
\label{fig:nD3.5_evol}  
\end{figure*}

\begin{figure*}
\begin{minipage}[b]{0.48\linewidth}
\epsfig{figure=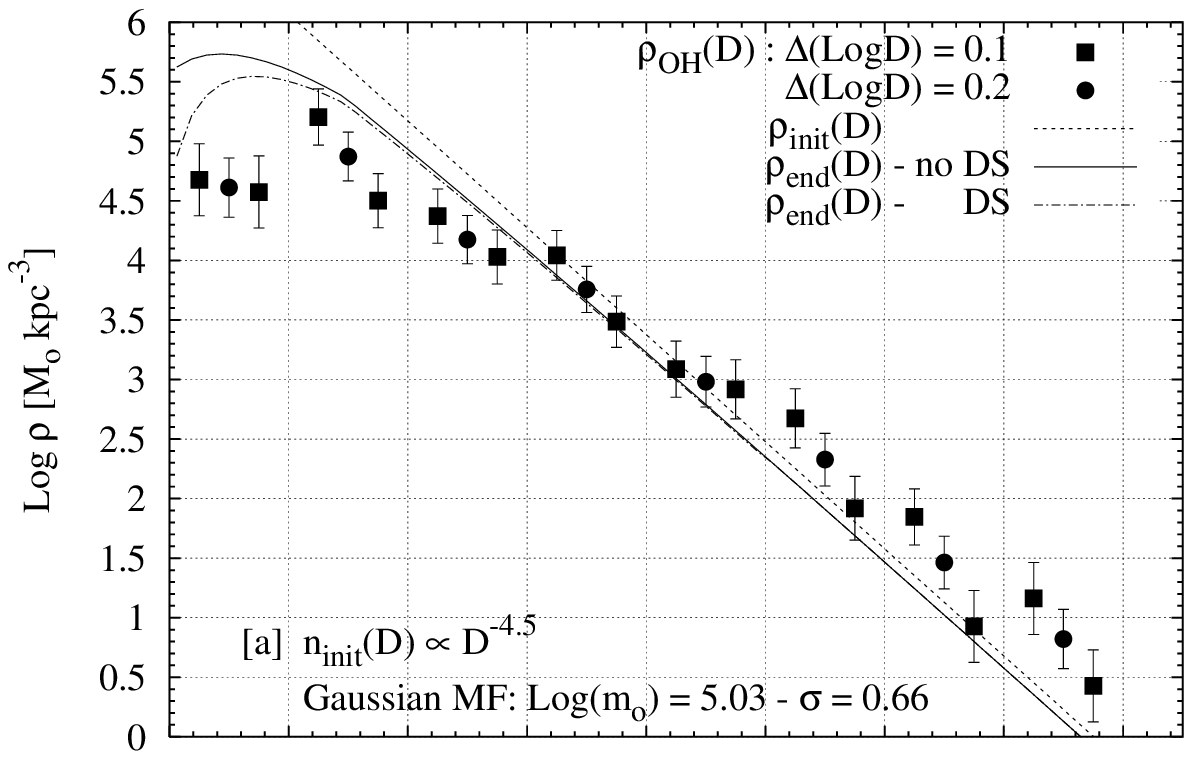, width=\linewidth}
\end{minipage}
\hfill
\begin{minipage}[b]{0.48\linewidth}
\epsfig{figure=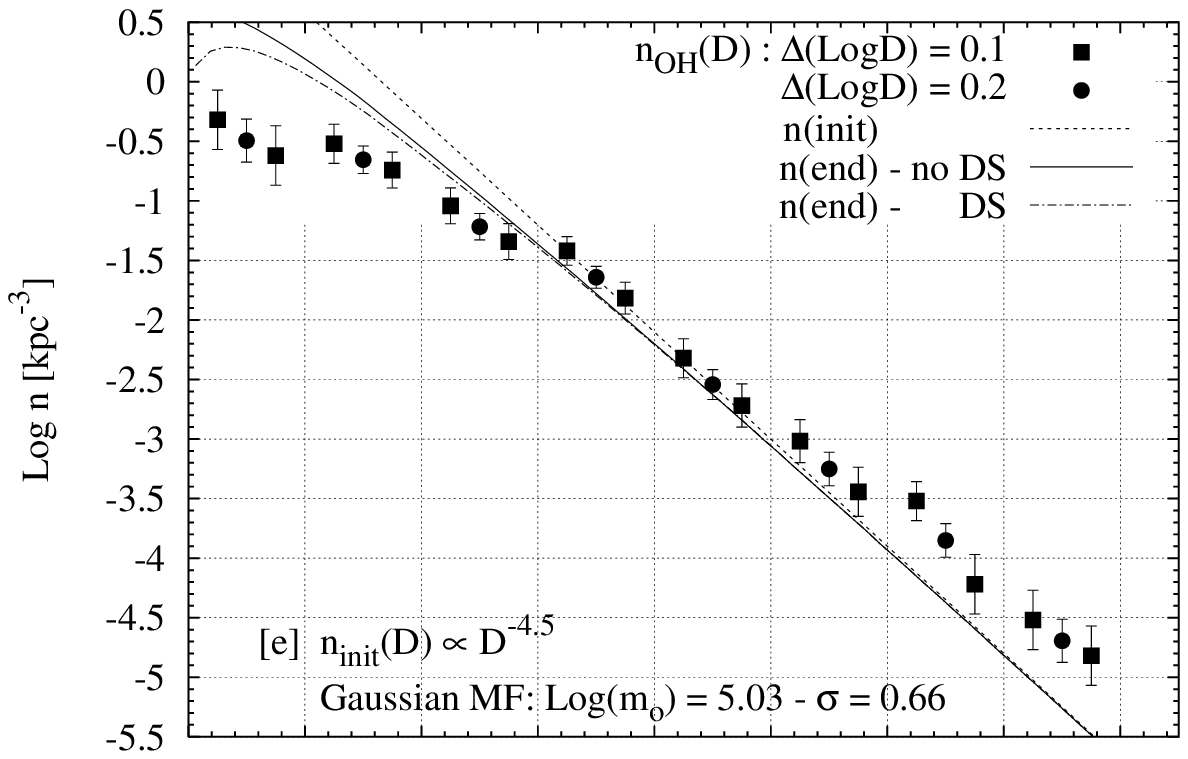, width=\linewidth}
\end{minipage}
\vfill
\vspace*{-8mm}
\begin{minipage}[b]{0.48\linewidth}
\epsfig{figure=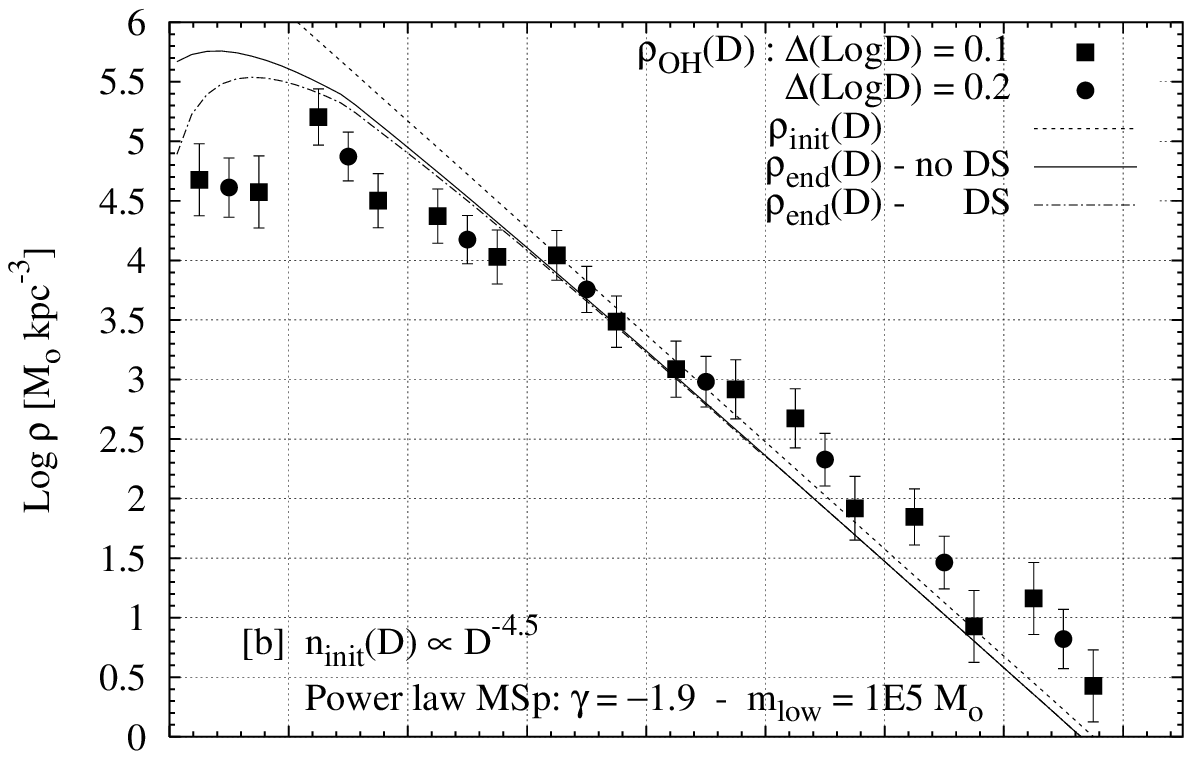, width=\linewidth}
\end{minipage}
\hfill
\begin{minipage}[b]{0.48\linewidth}
\epsfig{figure=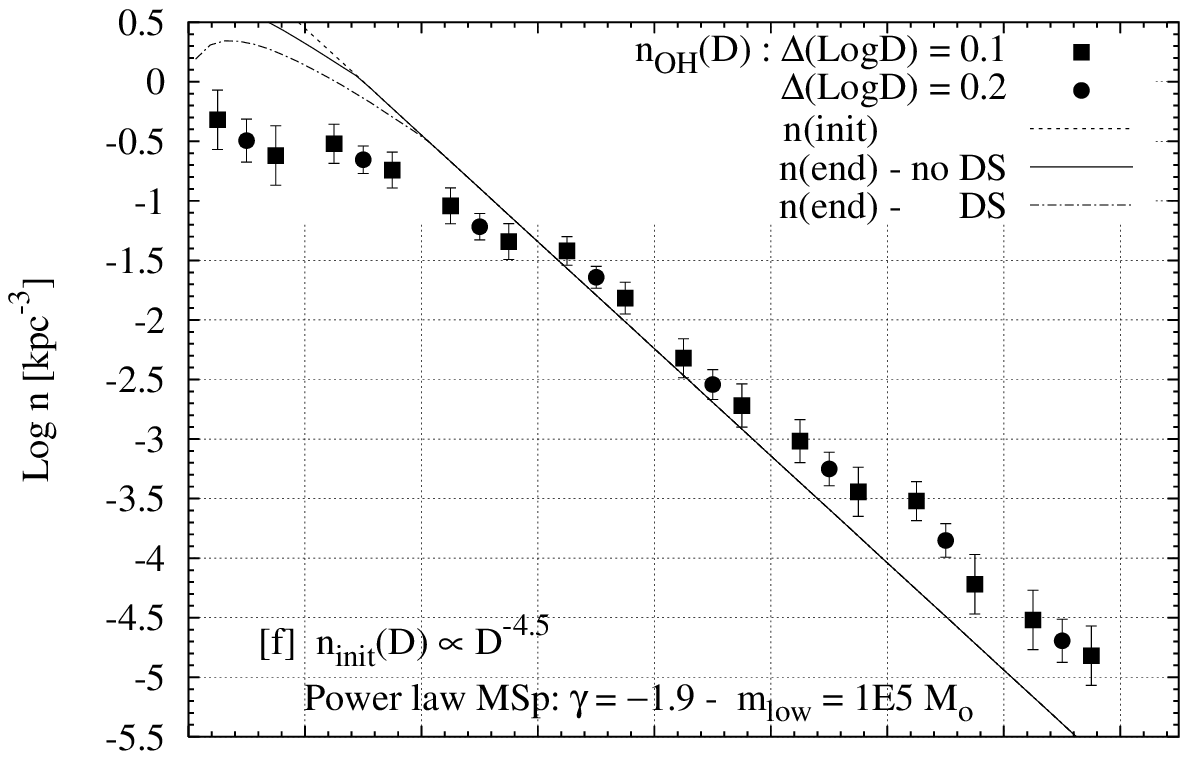, width=\linewidth}
\end{minipage}
\vfill
\vspace*{-8mm}
\begin{minipage}[b]{0.48\linewidth}
\epsfig{figure=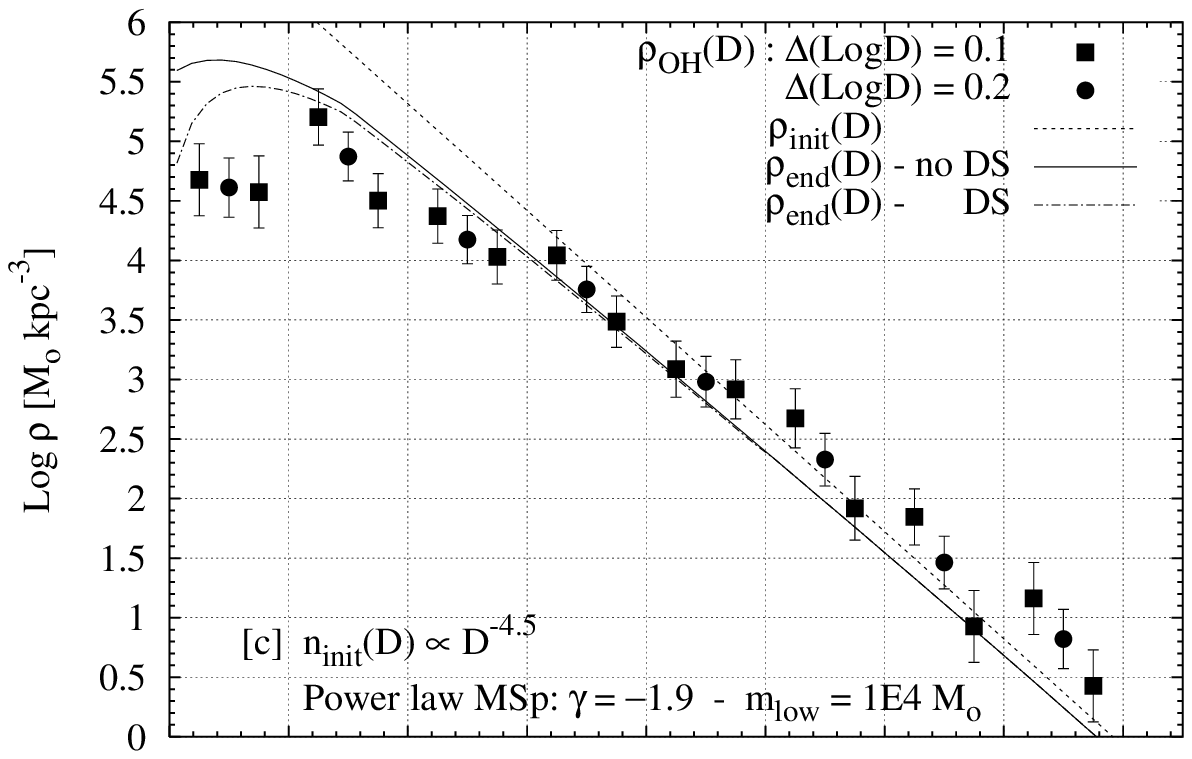, width=\linewidth}
\end{minipage}
\hfill
\begin{minipage}[b]{0.48\linewidth}
\epsfig{figure=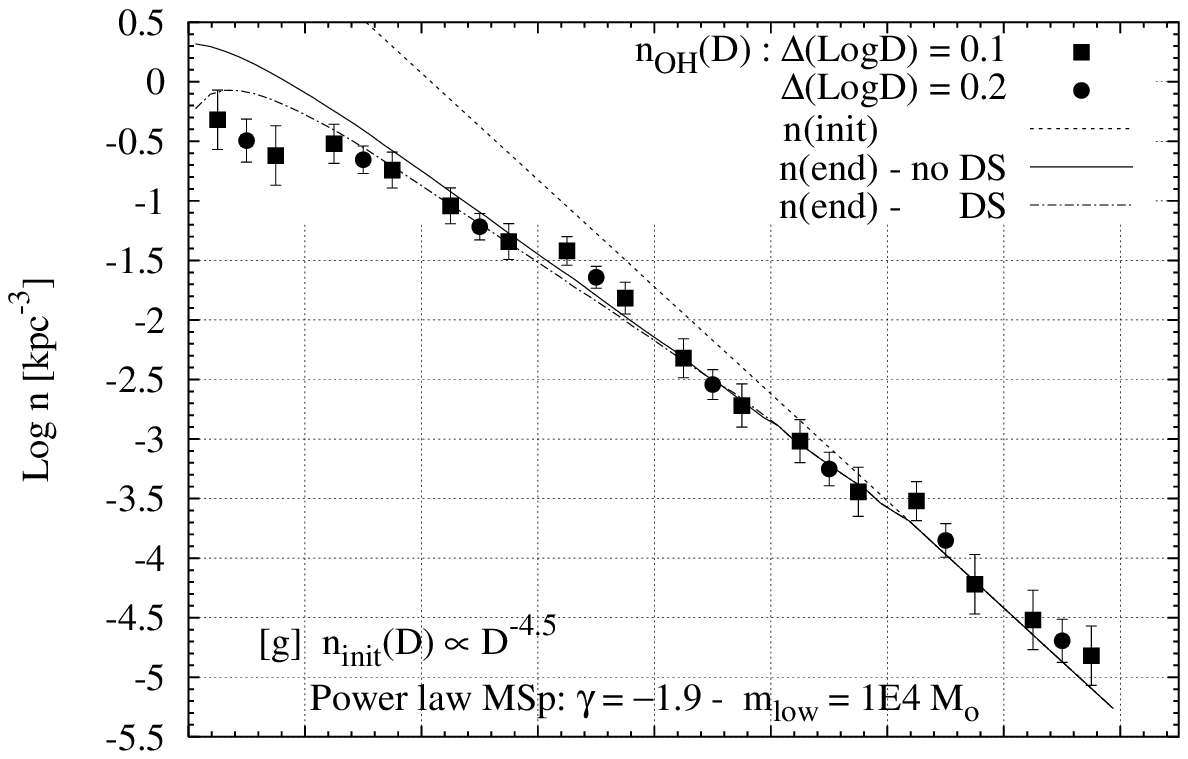, width=\linewidth}
\end{minipage}
\vfill
\vspace*{-8mm}
\begin{minipage}[b]{0.48\linewidth}
\epsfig{figure=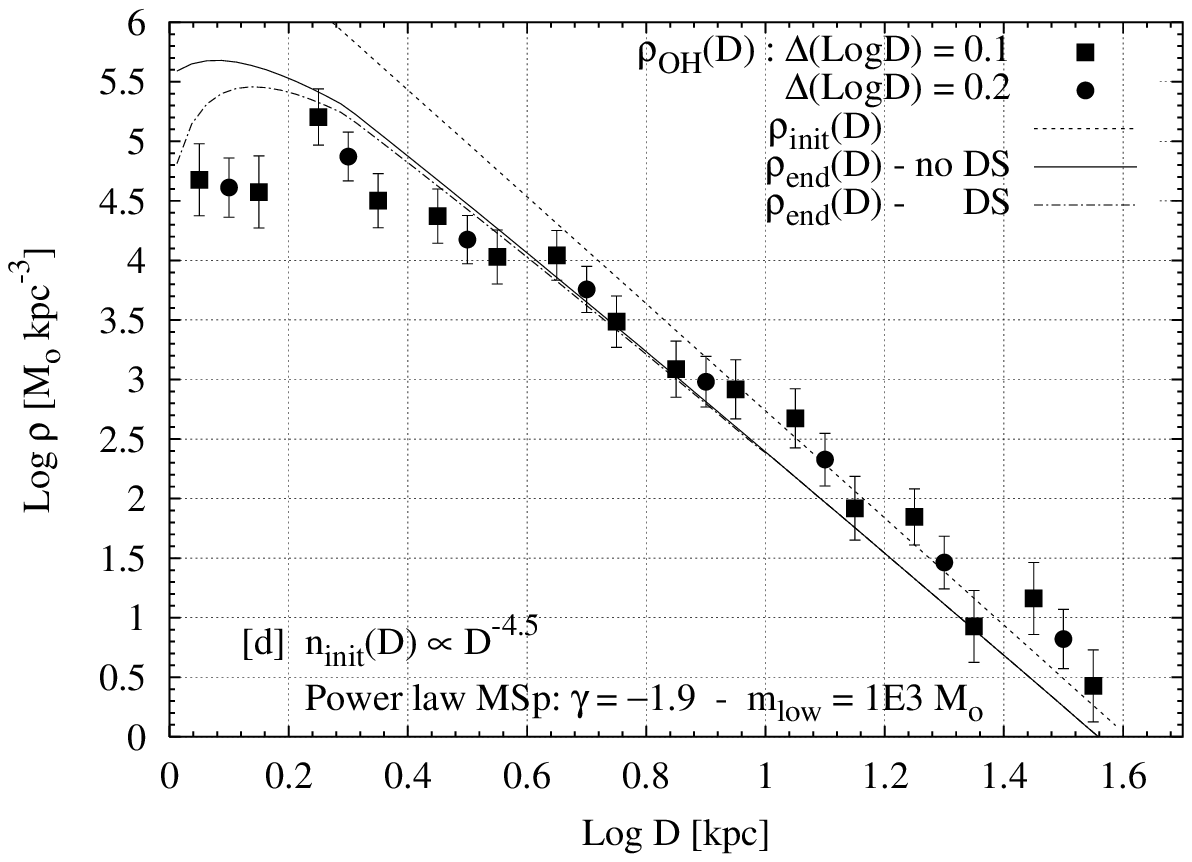, width=\linewidth}
\end{minipage}
\hfill \hspace*{5mm}
\begin{minipage}[b]{0.48\linewidth}
\epsfig{figure=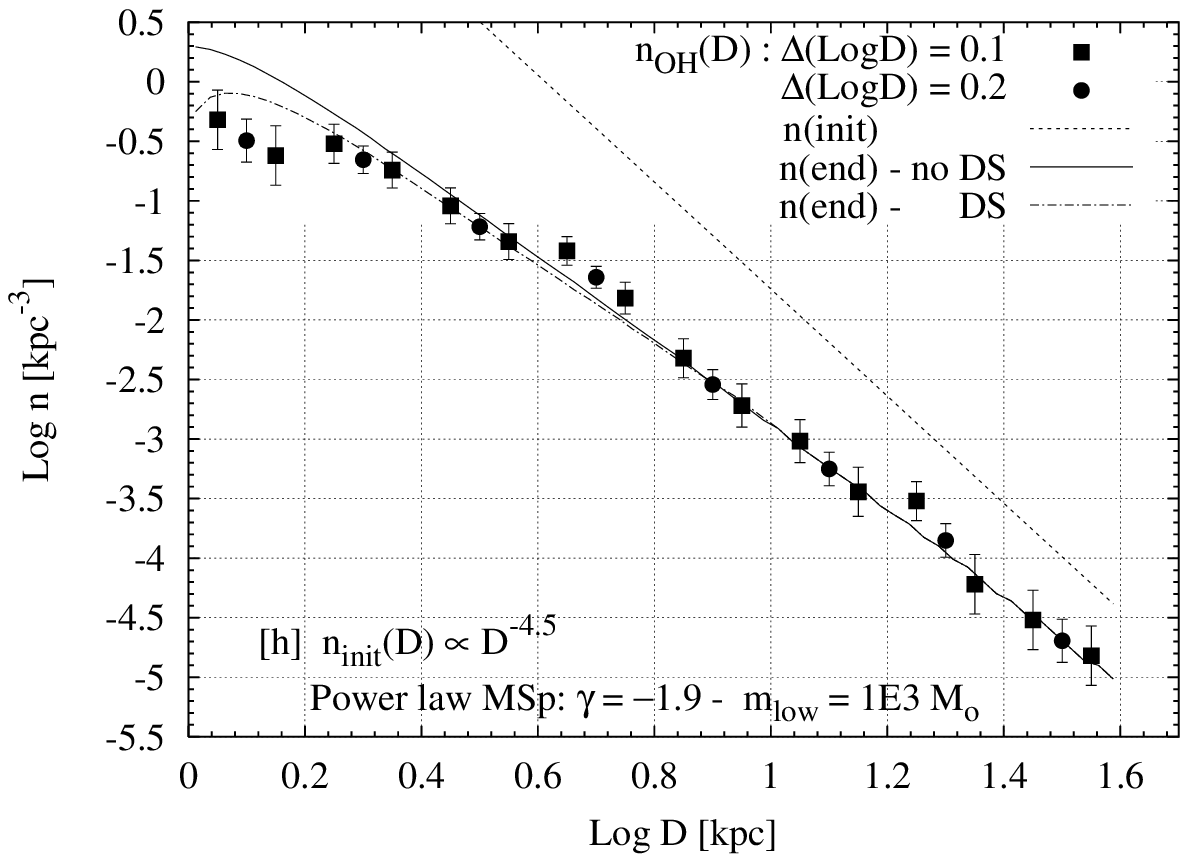, width=\linewidth}
\end{minipage} 
\caption{Same as in Fig.~\ref{fig:nD3.5_evol} but the initial
space-density obeys $n_{init} \propto D^{-4.5}$ }  
\label{fig:nD4.5_evol}  
\end{figure*}

The initial (dotted curves) and evolved (solid curves: no disc
shocking, dashed-dotted curves: with disc shocking) mass and number density 
profiles are displayed in the left and right panels, respectively, of
Figs.~\ref{fig:nD3.5_evol} ($n_{init} \propto D^{-3.5}$) and
\ref{fig:nD4.5_evol} ($n_{init} \propto D^{-4.5}$).  
Examination of the left panels shows that the mass density profile $\rho
(D)$ has remained fairly well preserved during evolution for
a Hubble-time \footnote{It should be kept
in mind that Vesperini \& Heggie 's (1997) model includes a gaseous mass loss
of 18 per cent fed by stellar winds.  This decrease in the cluster
mass is independent of galactocentric distance and, therefore, does
not alter the shape of the mass density profile even though this one
is reduced by the same amount.}.
The strongest change takes place for a power-law down to 
1000\,M$_{\odot}$.  Even in that case however, the slope of the
initial distribution is reduced by $\simeq 0.3$ only.
The initial steepness of the cluster system mass density profile is
thus robust, irrespective of the initial mass spectrum.  
In contrast, the evolution with time of the number density profile is
much dependent on initial conditions, as revealed by the right panels of 
Figs.~\ref{fig:nD3.5_evol} and \ref{fig:nD4.5_evol}.  While the
initial and evolved number density profiles are similar in case
of a gaussian mass function (i.e., an initial mass spectrum depleted in
low-mass clusters with respect to a power-law extending down to
low-mass objects) or in case of a power-law mass spectrum truncated
at 10$^5$\,M$_{\odot}$, the slope of the profile is reduced by $\simeq$ 1
if the initial power-law mass spectrum goes down to 1000\,M$_{\odot}$.     
Equation \ref{eq:mgc_t_noDS} shows that no globular clusters with
masses lower than 4000\,M$_{\odot}$ are able to survive within the
outer bound of the Old Halo (say, $D\lesssim 40$\,kpc).  Although
these clusters represent a small fraction of the cluster system mass
($\simeq 15$ per cent), by number, they account for  
$\simeq 70$ per cent of the clusters.  Hence, the destruction of these
low-mass objects leads to a significantly altered number density
profile while leaving the mass profile almost unaffected. 
Obviously, mass-related quantities are more reliable
indicators of the initial conditions than are number-related quantities,
especially in the case of an initial mass spectrum favouring
low-mass clusters.  In addition to possible slope alterations, in all
cases, the spatial distribution develops a core, that is, it gets
flattened in the galactocentric range 1 to 2\,kpc,
illustrating the larger efficiency of the disruptive processes closer
to the Galactic centre.  

\section{What density profiles tell us about the globular cluster 
initial mass function}
\label{sec:comp_mod_obs}

In Figs.~\ref{fig:nD3.5_evol} and \ref{fig:nD4.5_evol}, the evolved
spatial distributions (solid and dashed-dotted curves) have been
vertically shifted in order to provide the best least-squares
agreement with the presently observed distributions.  The initial
space-densities (dotted curves) have then been shifted accordingly.  
Table \ref{tab:fit_evol_GCS} shows the corresponding $\chi ^2$ and 
incomplete gamma functions $Q(\nu /2, \chi ^2/2)$ (i.e., the
goodness-of-fit).  \\

Considering the case of an initial spatial
distribution scaling as D$^{-3.5}$, we note the excellent agreement
between the evolved and observed mass density profiles, irrespective
of the initial cluster mass spectrum.  This is an expected result since 
the Old Halo mass density profile also scales as D$^{-3.5}$ 
(see Table \ref{tab:fit_pure_pl}) and since mass profiles are fairly
well-preserved in spite of the destruction of a significant fraction of
the initial cluster population.  Including disc shocking leads to a larger
decrease in mass density at small galactocentric distance and to an
even better modelling of the Old Halo profile.  We note in passing
that some values of the significance as measured by
the incomplete gamma function are puzzlingly high, that is,
larger than 0.8.  It is worth keeping in mind however that the initial spatial
distribution of relevance here (i.e., $n_{init} \propto D^{-3.5}$) is
not that much a model but, instead, derives from a fit to the observed 
mass density profile for D$\gtrsim 3$\,kpc (see Table
\ref{tab:fit_pure_pl}).  These high $Q$ values thus illustrate 
furthermore that the evolved mass density profile mirrors
faithfully the initial one.  

As for the evolved number density profile, this agrees with the data
only if the globular cluster system started with either a gaussian
mass function or a
power-law mass spectrum truncated at 10$^5$\,M$_{\odot}$.  In case of a 
power-law mass spectrum probing down to low cluster mass, the
$Q$ values get extremely low, disproving such initial mass spectra.  
This discrepancy results from the sharp change experienced by the
slope of the number density profile (panel [h] in
Fig.~\ref{fig:nD3.5_evol}), leading to a present steepness 
significantly shallower ($\simeq -2.5$) than the observed one ($\simeq -3.5$,
see Table \ref{tab:fit_pure_pl}).  
As a result, examination of Fig.~\ref{fig:nD3.5_evol} and Table
\ref{tab:fit_evol_GCS} shows that the comparison of the evolved (i.e.,
modelled) and observed radial density profiles, {\it in terms of mass
as well as in terms of number}, enables us to constrain the initial
mass spectrum of globular clusters.  The $Q$ values strongly favour 
either an initial gaussian mass function or a
power-law with slope of order $-1.9$ and truncated at large mass,
around 10$^5$\,M$_{\odot}$, that is, an initial mass distribution
which is somehow depleted in low-mass objects.  
In this respect, our results confirm those achieved by Vesperini~(1998). \\
     
We now consider the steeper initial spatial distribution,
namely $n_{init} \propto D^{-4.5}$.  As already suggested by Baumgardt
(1998), there is an excellent agreement between the halo number
density profile and its modelled counterpart in the case of a power-law
extending down to 1000\,M$_{\odot}$, especially  
if disc shocking is taken into account.  Owing to the large contribution
of low-mass clusters, the initially steep profile is turned into a
shallower one, thus matching the halo $-3.5$ slope.  
However, we caution that the evolved mass density profile does not
fit its Old Halo counterpart convincingly in any case.  The best match is
obtained for a globular cluster system with a power-law initial mass 
spectrum probing down to low-mass (see Table \ref{tab:fit_evol_GCS}).        
Even though such a possibility cannot be ruled out firmly, the goodness-of-fit
is very marginal ($Q \lesssim0.001$).  This case is therefore much less
likely than the one we have previously discussed, namely, a globular 
cluster system whose initial spatial steepness is similar to the present one.  
     
Baumgardt (1998) himself noted the oddity of this result as it implies
a discrepancy between the initial slope of the globular cluster spatial
distribution on the one hand and the steepness of the stellar halo
density profile on the other hand.  Indeed, the space-density of halo
RR Lyrae (Suntzeff, Kinman \& Kraft 1991) as well as of halo blue
horizontal branch stars (Kinman, Suntzeff \& Kraft 1994) falls off as
$D^{-3.5}$.  Baumgardt (1998) suggests that this discrepancy results
from a varying star cluster to   
field star formation efficiency.  Considering the $Q$ values listed in
Table \ref{tab:fit_evol_GCS}, a much safer conclusion may be that the 
globular cluster system started with a space-density scaling as 
$D^{-3.5}$ coupled with either a gaussian mass function or a power-law 
truncated at 10$^5$\,M$_{\odot}$. \\

While our simulations start with many thousands of clusters, the
survival rates $F_N$ quoted in Table \ref{tab:frac_surv} indicate 
that the initial number of clusters is on the order of that today in case of
a gaussian initial mass function or in case of a power-law mass spectrum
truncated at $10^5$\,M$_{\odot}$.  As for the gaussian initial mass
function, we have checked that an initial total number of 
clusters of 200 only does not introduce a significant scatter in the
evolved mass density profile with respect to the size of the  error bars. 
Using Vesperini \& Heggie 's (1997) model with disc shocking and
considering a slope $\gamma = -3.5$ for the initial radial
distribution, the incomplete gamma function for the evolved mass
density profile ($\Delta log D = 0.2$) ranges from 0.3 up to 0.9 (10
random samplings of the 
gaussian cluster IMF).  We note that these results are obtained in
case of a gaussian truncated at 1.5E6\,M$_{\odot}$ in order to avoid
the presence of clusters significantly more massive than is observed
today.  Actually, inspection of the luminous mass estimates of the
halo globular clusters shows 
$10^6$\,M$_{\odot}$ to be an upper limit to the present-day globular
cluster mass.  More massive clusters do actually exist, e.g., 
$\omega$ Cen, M54 (=Sagittarius core), NGC 2419 and a few disc
clusters, but none of them are relevant to the present study.  
Running the same simulations in case of a
non-truncated gaussian, the incomplete gamma function tends to get
smaller (i.e., down to 0.002 in one case).  This is due to the
occasional sampling of very massive (i.e.,$>$ 1.5E6\,M$_{\odot}$) clusters,
giving rise to an upwards scatter in the outermost least-populated 
bins of the mass density profile. 

As for the case of a power-law mass spectrum 
truncated at $10^5$\,M$_{\odot}$, we have evolved a cluster system
initially comprising 150 clusters only.  The incomplete gamma function
for the evolved mass density profile ranges from 0.01 to 0.7 (10
random realisations, of which 8 give $Q > 0.1$). 
This robustness, despite a limited number of clusters, is due to the
narrow mass range associated to the truncation at large cluster mass. 
\footnote{In case of a power-law initial mass
spectrum extending down to
1000\,M$_{\odot}$, the globular cluster system initially hosted
several thousands of clusters and the small-number sampling thus does
not constitute an issue.  Should the initial number of clusters be
on the order of that today (say, 100), the evolved 
population would contain just a very few survivors, possibly none at
all since the selection of low-mass clusters would be favoured owing to the
small-number sampling.  Actually, following a Hubble-time of
evolution, the number of survivors is in the range [2,10] ([0,4]) if
the slope of the initial radial distribution is $-3.5$ ($-4.5$) (10 random
realisations), leading to a discrepancy with the present-day cluster
number.}  \\

In fact, our results are reminiscent of 
those obtained by Vesperini (2000, 2001) in the case of globular 
cluster systems hosted by elliptical galaxies.  Investigating the 
case of a power-law initial mass spectrum extending down to low-mass 
combined with a coreless $D^{-3.5}$ initial number density profile, 
Vesperini (2001) noted that the evolutionary 
processes produce a significant dependence of the average cluster mass
on the galactocentric distance in the sense that clusters 
located in the inner galactic regions are more massive.  This
dependence is equivalent to the difference between the slopes of the
evolved mass and number density profiles highlighted in the bottom
panels of Fig.~2.  Vesperini 's (2001) result contrasts with
several observational studies that fail to find a significant radial
gradient of the average cluster mass within elliptical galaxies (e.g.,
M87, Kundu et al.~1999).  Conversely, Vesperini (2000) emphasized that,
in the case of a gaussian initial mass function, the radial gradient
of the average cluster mass is weak and consistent with the observations.
Equivalently, the evolved mass and number density profiles match each
other, as shown by the top panels of Fig.~2.

The marked evolution of the number density profile with respect to the
mass density profile in the case of a power-law mass spectrum arises
because, as Vesperini (1998, 2000, 2001), we have assumed that
clusters are orbiting at constant galactocentric distance.  
Should a substantial radial mixing take place, a more radially
uniform mass function may emerge.  We have thus tested whether our
results are significantly affected if the orbital 
eccentricity is $e=0.5$. According to Baumgardt \& Makino (2003), the
lifetime of a cluster on an orbit with eccentricity $e$ is decreased by
a factor $(1-e)$ with respect to a cluster on a circular orbit with a
radius similar to the apogalactic radius of the eccentric orbit.
Considering the two extreme (and key) cases, namely, the gaussian [G] and
power-law [PL3] initial cluster mass distributions, combined with 
a coreless $D^{-3.5}$ initial number density profile, we have run additional
simulations in which the quantity $F_{cw}$ (equation \ref{eq:Fcw}) is 
halved (i.e., multiplied by $1-e$).  The corresponding evolved mass
and number density profiles are in remarkable agreement with those 
derived under the assumption of circular orbits.  
The only significant difference is the
destruction of all the clusters confined within 1.2\,kpc from the
Galactic centre if $e=0.5$.  The comparison between the predicted profiles 
and the Old Halo distributions leads to (considering $\Delta logD
= 0.2$): 
\begin{enumerate}
\item for a Gaussian initial mass function [G]: Q=0.03 [$\rho (D)$]
and Q=0.003 [$n (D)$] for the mass and number density profiles, respectively;
\item for a power-law initial mass spectrum [PL3]: Q=0.005 [$\rho (D)$] and
Q=10$^{-16}$ [$n (D)$]. 
\end{enumerate}
As for the gaussian mass function, the incomplete gamma
function is smaller than that derived for circular orbits 
(Table 4).  Yet, this effect is mostly driven by the first bin, this
being located on the edge of the region of complete cluster
destruction (i.e., D $\simeq$ 1.2kpc).  We note that the
power-law initial mass spectrum [PL3] is again rejected by the poor 
agreement between the predicted and observed number density profiles
(that is, the increase in the mean cluster mass with decreasing
galactocentric distance is much larger than is observed).  This result
is not unexpected.  The evaporation rate of a cluster with a given
mass depends on its orbit, especially its pericentre (see e.g., 
Baumgardt 1998).  
Considering a system of clusters extending up to 40\,kpc from the
Galactic centre, if e=0.5, the range of perigalactic distances is 
$\simeq$ 1-13.3 kpc.  The lack of consistency between the slopes of
the evolved mass and number density profiles in the [PL3] case
illustrates that such a range of perigalactic distances is still too
large to erase radial variations in the 
mean cluster mass.  As demonstrated by Fall \& Zhang (2001), 
a narrow distribution of pericentres and, thus, disruption rates
almost independent of the cluster galactocentric distance,
can be achieved if the initial velocity distribution shows some 
radial anisotropy {\it and} the radial anisotropy is increasing outward.
Should such conditions be satisfied, the mean cluster masses in the
inner ($D < 5$\,kpc)  and outer ($D >5$\,kpc) groups of clusters are
similar. 
However, that result is obtained in the case of an initial Schechter 
mass spectrum (i.e., $dN/dm \propto m^{-2} e^{-m/5 \times
10^6\,M_{\odot}}$), i.e., a mass distribution steeper than 
a power-law $\alpha = -2$.  In the case of an initial power-law mass
spectrum, the evolved mass distribution fails to reproduce the high 
mass-regime of the present-day distribution (Fall \& Zhang 2001, 
their Fig.~3).  
A radially uniform cluster mass function can thus be achieved 
even if the initial mass distribution is steadily rising toward low-mass, 
although such a solution requires appropriate tuning.

\section{Summary and conclusions}
The initial distribution in mass of the Galactic halo globular
clusters has so far remained a poorly-known function.  
This is due to the fact that both a gaussian initial mass function 
(i.e., a two-index power-law mass spectrum) and a power-law initial 
mass spectrum evolve into the presently observed bell-shaped cluster 
mass function.  As a result, the study of the temporal
evolution of the mass function/spectrum only does not enable one to assess
how large was the contribution of low-mass (say a few
thousand solar masses) objects to the initial population of clusters.
  
In this paper, we have proposed a new method for shedding light on
this issue which consists in comparing globular cluster system density 
profiles with their modelled counterparts, as a function of mass
density as well as a function of number density.  We assume, as in
previous studies of globular 
cluster system dynamical evolution, that the cluster mass range and
the cluster mass spectrum are initially independent of their 
galactocentric distance.
This is equivalent to assuming that the initial mass and number 
density profiles are identical in shape.  On the other hand, the
present mass and number density profiles of the Old Halo are alike as
well (see Section 2).  The new constraints on the initial
globular cluster mass function we have derived arise from 
combining this widespread assumption and this observational fact.  

The clusters most vulnerable to evaporation and disruption are the low-mass
ones as well as those located at small galactocentric distance.  
In other words, the disruption rate of globular clusters depends 
on their initial distribution in space as well as on their initial 
distribution in mass.
The key point is that the mass and number density profiles show
contrasting temporal evolutions.  While the evolution of the latter is
heavily driven by the initial cluster mass spectrum, the former
remains almost unchanged during evolution for a Hubble-time, providing
always that the Galactic potential remains smooth and was slowly
varying.  Hence, the robustness of the mass density profile provides
us with an immediate estimate of the initial steepness of the spatial
distribution.  This can then be evolved for various initial cluster
mass spectra and the resulting number density profiles compared to the
observed one in order to discriminate which cluster mass spectrum
provides the best match between the data and the model.

In order to test this idea, we have evolved various putative globular
cluster systems characterized 
by different combinations of initial number density profiles
(i.e., how clusters are distributed in space) and initial mass spectra
(i.e., how clusters  are distributed in mass).  Results of these simulations
are displayed in Figs.~\ref{fig:nD3.5_evol} and \ref{fig:nD4.5_evol}. 
We have shown that, irrespective of the initial globular cluster mass
spectrum,  
the damage performed to the initial mass content in clusters is limited
to one effective radius, that is, D$\lesssim 3$\,kpc (see also McLaughlin
1999).  While in this range, the spatial distribution of the
cluster system mass flattens owing to the
greater efficiency of cluster destruction processes, the overall slope
remains close to its initial value.  In sharp contrast, the temporal
evolution of the number density profile depends sensitively on the
initial mass spectrum.  The steepness of the space-density is
stationary in case of a gaussian mass function or of a power-law truncated at
10$^5$\,M$_{\odot}$.  On the other hand, it gets significantly
shallower in the case of a mass spectrum favouring low-mass clusters, 
e.g., a power-law extending down to 10$^3$\,M$_{\odot}$.
     
For each simulation, we have compared in a least-squares sense the
presently observed spatial distributions with the modelled ones,
obtaining the $\chi ^2$ and the incomplete gamma function measure
of probability (see Table \ref{tab:fit_evol_GCS}).
The most likely initial conditions of course correspond to 
the cases for which the evolved mass {\it and} number density
profiles are in good agreement with their Old Halo counterparts.
The best match is achieved when an initial spatial distribution with a
slope of $-3.5$ is combined with an initial mass spectrum depleted in
low-mass clusters, that is, either a gaussian mass function or a
power-law mass spectrum truncated at 10$^5$\,M$_{\odot}$.  
In this case, the cluster destruction rate is limited, as also is the
corresponding temporal evolution of the number density profile, thus
preserving its initial $-3.5$ steepness, in agreement with what is
observed for the Old Halo (see Table \ref{tab:fit_pure_pl}) .  If the
Galactic halo globular cluster system had actually started with this 
initial spatial distribution,  
it is very unlikely that the initial mass spectrum was a pure
power-law extending down to 1000\,M$_{\odot}$.  The abundance of
low-mass objects in such a globular cluster system would make the
number density profile shallow with time, making it unable to fit the 
present $-3.5$ slope. 

We confirm Baumgardt 's (1998) finding following which a power-law
probing down to 1000\,M$_{\odot}$ combined with a steep (i.e., $\gamma
= -4.5$) spatial distribution leads to good agreement with the
observed present-day number density profile.  However, we caution that the
observed mass density profile is then not well fitted by its evolved
counterpart. 
In fact, owing to their robustness, all the evolved mass density
profiles, irrespective of the initial globular cluster mass spectrum, 
are locked close to their initial $-4.5$ slope.  As a result, 
they remain significantly steeper, even after a 15\,Gyr long
evolution, than the $-3.5$ slope shown by the present spatial 
distribution of the Old Halo cluster system mass. 
Even though such a possibility cannot be firmly ruled out ($Q \lesssim
0.001$ if disc shocking is included in the simulations), it
remains much less likely than the initial $D^{-3.5}$ space-density
which we have just discussed.  As a result, although the number 
density profile alone indicates that the Galactic globular cluster system
may have started with a very steep initial spatial distribution and 
a power-law mass spectrum covering three orders of magnitude in mass, 
the mass density profile tends to dismiss this possibility. 
   
All together, our results support the hypothesis following which the
Galactic halo globular cluster system started with an initial 
space-density scaling as
D$^{-3.5}$ and an initial mass spectrum somehow depleted in low-mass
clusters, that is, a bell-shaped mass function similar to the current
one, or a power-law mass spectrum truncated near 10$^5$\,M$_{\odot}$.

\bsp

\section*{Acknowledgments}
This research was supported by a Marie Curie Intra-European
Fellowships within the $6^{th}$ European Community Framework Programme.

\label{lastpage}

\end{document}